\def\BibTeX{{\rm B\kern-.05em{\sc i\kern-.025em b}\kern-.08em
    T\kern-.1667em\lower.7ex\hbox{E}\kern-.125emX}}

\documentclass[journal]{IEEEtran}

\usepackage{amsmath, graphics,amssymb,epsfig,subfigure}
\usepackage{algorithm}
\usepackage{algorithmic}
\usepackage{float}
\usepackage{multirow}
\usepackage{cite}
\usepackage{multicol}

\newlength{\capwidth}
\newtheorem{Lemma}{Lemma}
\newtheorem{Proposition}{Proposition}

\newcommand{\yv}{\mbox{$\bf y $}}

\newcommand{\fv}{\mbox{$\bf f $}}
\newcommand{\Yv}{\mbox{$\bf Y $}}
\newcommand{\Hv}{\mbox{$\bf H $}}
\newcommand{\Dv}{\mbox{$\bf D $}}
\newcommand{\Av}{\mbox{$\bf A $}}
\newcommand{\xv}{\mbox{$\bf x $}}

\newcommand{\Sv}{\mbox{$\bf S $}}

\newcommand{\Pv}{\mbox{$\bf P $}}
\newcommand{\pv}{\mbox{$\bf p $}}
\newcommand{\nv}{\mbox{$\bf n $}}

\newcommand{\Cv}{\mbox{$\bf C $}}

\newcommand{\Mv}{\mbox{$\bf M $}}
\newcommand{\mv}{\mbox{$\bf m $}}
\newcommand{\Kv}{\mbox{$\bf K $}}
\newcommand{\kv}{\mbox{$\bf k $}}

\newcommand{\Bv}{\mbox{$\bf B $}}
\newcommand{\Qv}{\mbox{$\bf Q $}}

\newcommand{\Wv}{\mbox{$\bf W $}}
\newcommand{\Tv}{\mbox{$\bf T $}}

\newcommand{\Iv}{\mbox{$\bf I $}}

\newcommand{\Rv}{\mbox{$\bf R $}}

\newcommand{\Fv}{\mbox{$\bf F $}}

\newcommand{\Lv}{\mbox{$\bf L $}}
\newcommand{\Gv}{\mbox{$\bf G $}}

\newcommand{\uv}{\mbox{$\bf u $}}

\newcommand{\Uv}{\mbox{$\bf U $}}
\newcommand{\Vv}{\mbox{$\bf V $}}
\newcommand{\Zv}{\mbox{$\bf Z $}}
\newcommand{\vv}{\mbox{$\bf v $}}

\newcommand{\be}{\begin{equation}}

\newcommand{\ee}{\end{equation}}
\newcommand{\bea}{\begin{eqnarray}}
\newcommand{\eea}{\end{eqnarray}}
\newcommand{\bdp}{\begin{displaymath}}
\newcommand{\edp}{\end{displaymath}}

\renewcommand{\det}{{\hbox{det}}}
\newcommand{\trace}{{\sf Tr}}
\newcommand{\transp}{{\sf T}}

\begin{document}
\title{Optimal Precoder Designs for Sum-utility Maximization in SWIPT-enabled Multi-user MIMO Cognitive Radio Networks}

\author{\IEEEauthorblockN{Changick Song, \emph{Member, IEEE}, Hoon Lee, \emph{Member, IEEE}, and Kyoung-Jae Lee, \emph{Member, IEEE}}\\
\thanks{C. Song is with the Department of Electrical Engineering, Korea National University of Transportation, Chungju, Korea, 27469 (e-mail: c.song@ut.ac.kr).

H. Lee is with the Information Systems Technology and Design Pillar, Singapore University of Technology and Design, Singapore 487372 (e-mail: hoon\_lee@sutd.edu.sg).

K.-J. Lee is with the Department of Electronics and Control Engineering, Hanbat National University, Daejeon, Korea, 34158 (e-mail: kyoungjae@hanbat.ac.kr).
}
\vspace{-15pt} } \maketitle

\begin{abstract}
In this paper, we propose a generalized framework that combines the cognitive radio (CR) techniques
for spectrum sharing and the simultaneous wireless information and power transfer (SWIPT) for energy harvesting (EH)
in the conventional multi-user MIMO (MuMIMO) channels, which leads to an MuMIMO-CR-SWIPT network.
In this system, we have one secondary base-station (S-BS) that supports multiple secondary information decoding (S-ID) and secondary EH (S-EH) users
simultaneously under the condition that interference power that affects the primary ID (P-ID) receivers should stay below a certain threshold.
The goal of the paper is to develop a generalized precoder design that maximizes the sum-utility cost function
under the transmit power constraint at the S-BS, and the EH constraint at each S-EH user, and the interference power constraint at each P-ID user.
Therefore, the previous studies for the CR and SWIPT systems are casted as particular solutions of the proposed framework.
The problem is inherently non-convex and even the weighted minimum mean squared error (WMMSE)
transformation does not resolve the non-convexity of the original problem.
To tackle the problem, we find a solution from the dual optimization via sub-gradient ellipsoid method
based on the observation that the WMMSE transformation raises zero-duality gap between the primal and the dual problems.
We also propose a simplified algorithm for the case of a single S-ID user, which is shown to achieve the global optimum.
Finally, we demonstrate the optimality and efficiency of the proposed algorithms through numerical simulation results.
\end{abstract}
\section{Introduction}\label{sec:introduction}

Recently, cognitive radio (CR) technologies have been developed as a promising solution for efficient spectrum usage.
It was shown that even when the licensed primary users are active for transmission or reception,
the unlicensed secondary users are still able to share the spectrum opportunistically
with the active primary users by utilizing multiple transmit antennas and properly designing its transmit spatial spectrum.
The fundamental limit of such a network was studied in \cite{MGastpar:07}.
More feasible approaches have also been discussed in \cite{RZhang:08,RZhang:10,KyoungJae:11b} and references therein
to provide linear precoders
that maximize the weighted sum-rate or minimize the minimum mean squared error (MMSE) of the secondary users
by imposing constraints on the interference power at the primary receivers.

In the meantime, the idea of energy harvesting (EH) has recently been introduced to
provide convenient and sustainable energy supplies.
In particular, considering the information carrying radio frequency (RF) signals as a new energy source for the EH,
simultaneous wireless information and power transfer (SWIPT) techniques have garnered a lot of interest.
Recently, new advances in hardware technologies have enabled power to be transferred and harvested efficiently over a distance
\cite{XLu:15} \cite{SBi:15}.
However, appropriate precoder designs  based on multi-input multi-output (MIMO) antennas
are still essential to fully exploit the advantages of SWIPT networks
by concurrently maximizing the spectral efficiency of the information decoding (ID) users and
the amount of harvested energy at the EH users.
From this viewpoint, various precoding techniques have been investigated in multi-user SWIPT environments
\cite{RZhang:13,CSong:14GCW,CSong:16TWC,CSong:18TVT,JXu:14,JRubio:17,ZZhu:16,ZZhu:17,ZZhu:18,HLee:17CL,ZZong:16}.

The authors in \cite{RZhang:13} considered a two-user broadcasting channel (a single ID and a single EH)
in terms of maximizing the information rate to the ID user under a single EH constraint.
The result was then re-interpreted in \cite{CSong:14GCW,CSong:16TWC,CSong:18TVT} with respect to
the weighted MMSE (WMMSE) criterion, and more generalized and efficient solutions were provided.
The work in \cite{JXu:14} solved a transmit power minimization problem
under multiple signal-to-noise ratio (SNR) and EH constraints for the ID and EH users, respectively.
However, all the users were restricted to having a single antenna.
To take into account the general multi-user MIMO (MuMIMO) SWIPT environment where all ID and EH users are equipped with multiple antennas,
\cite{JRubio:17} proposed a precoder design based on the multi-objective cost function
to overcome the non-convex problem of the transmit covariance matrices in multi-stream MuMIMO SWIPT networks.
However, high computational complexity is still an issue,
because  a string of semi-definite programming (SDP) problems should be solved for each filter update during the iterative algorithm.
In \cite{ZZhu:16,ZZhu:17,ZZhu:18}, the security issue has been addressed for the beamforming designs in the multi-antenna SWIPT networks.

In the SWIPT networks, it is
often required to achieve the high received signal power to satisfy the energy requirement of each EH user,
which may also incur strong interference to other nearby users and networks that utilize the same spectrum.
Therefore, a practical SWIPT system should come with a proper interference management technique.
To address such an issue, an efficient beamforming scheme has been developed in \cite{FZhu:14}
to control the interference power to the primary networks in the CR-SWIPT topology.
However, the study was limited to a single antenna single ID user environment.



\begin{figure}
\begin{center}
\includegraphics[width=2.8in]{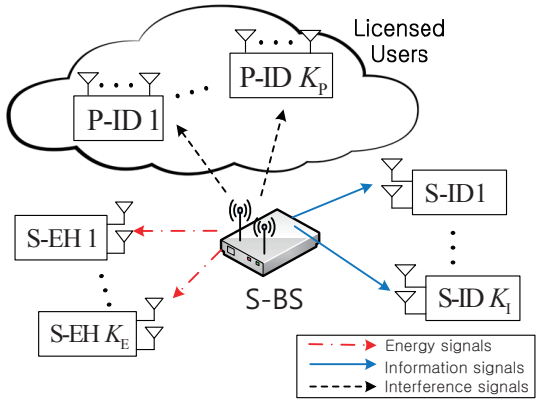}
\end{center}
\caption{System model for the proposed MuMIMO-CR-SWIPT networks \label{figure:SystemModel.eps}}
\end{figure}

In this paper, we investigate the optimal precoder designs in a general {\it MuMIMO CR SWIPT} network,
where one secondary base-station (S-BS) supports multiple secondary ID (S-ID) and multiple secondary EH (S-EH) users
all having multiple antennas by utilizing the licensed spectrum assigned to the primary ID (P-ID) users as depicted in Fig. \ref{figure:SystemModel.eps}.
We model our transmitter design as a unified framework for sum-utility maximization in the MuMIMO-CR-SWIPT systems,
in which the previous studies in \cite{RZhang:13,CSong:14GCW,CSong:16TWC,JXu:14,JRubio:17,FZhu:14} are shown to
be particular solutions of the proposed framework. The sum-utility maximization problem has initially been proposed in \cite{QShi:11}
to address different types of cost functions such as the weighted sum-rate (WSR), proportional fairness (PF),
and harmonic mean rate (HMR) at once in the conventional MuMIMO systems.
However, it remains unclear whether such existing solutions are applicable to the general MuMIMO-CR-SWIPT networks,
because the WMMSE transformation techniques in \cite{QShi:11} no longer resolve the non-convexity of the original problem even with respect to the precoding matrix only.

A main difference point of our study from the aforementioned previous works is that
we treat more general sum-utility maximization problem for the MuMIMO-CR-SWIPT systems,
for which the transmit power constraint at the S-BS, the interference power constraint at each P-ID user, and the EH constraint at each EH-user
must be satisfied simultaneously.
Although the previous work in \cite{FZhu:14} also considers the CR-SWIPT topology,
it is confined to the single S-ID user and a single receive antenna scenarios, which results in an easy-to-solve convex problem.
In contrast, our  problem is generally non-convex, and thus more challenging to solve.
Note that the conventional SDP approach as in \cite{JRubio:17} is not directly applied to our problem due to the non-linear utility cost functions.
To tackle the problem, we demonstrate that the WMMSE problem transformation gives a zero duality gap between the primal and its dual problems.
Then, we find the optimal solution by solving the dual problem that is given in the form of non-differentiable convex functions via the sub-gradient ellipsoid method.

The contribution of the paper is summarized as follows.
\begin{itemize}
  \item In Section \ref{sec:Maximum Achievable Energy}, we propose an optimal energy transmission scheme to maximize the amount of energy harvested at the S-EH users under the CR constraints.
  We first show that the rank-$1$ transmission is optimal for maximizing the weighted sum of harvested energy
  regardless of the number of EH and CR constraints.
  Then, we propose an efficient algorithm to find the optimal energy beam vector based on the subgradient ellipsoid method.
  This scheme enables us to identify a feasible range of the energy thresholds in the subsequent precoder designs.

  \item In Section \ref{sec:Sum-Utility Maximization}, we propose an optimal precoder design for general sum-utility maximization in the MuMIMO-CR-SWIPT networks.
  First, we prove that the WMMSE problem transformation gives rise to a strong duality in terms of the precoding matrix.
  Then, we propose an efficient algorithm to find the optimal precoder by adopting the ellipsoid and alternating optimization methods in its inner and outer iterations, respectively. The proposed algorithm converges at least to a locally optimal point,
  and thus can be made arbitrarily close to the global optimum with the aid of multiple initial points.
  A modified algorithm is also introduced to address the zero-interference constraints to the primary users.

  \item In Section \ref{sec:Joint optimal design}, we provide a simplified algorithm that achieves a globally optimal solution considering a special case of
  the S-BS supporting one S-ID user at a time in a time division multiple access (TDMA) manner,
  which is called {\it single user MIMO (SuMIMO) CR SWIPT}.
  Unfortunately, the problem is still non-convex. To resolve the problem, we first identify an optimal precoder structure as a closed-form through the Lagrange dual analysis.
  Then, we propose an efficient algorithm to determine the remaining dual variables based on the subgradient ellipsoid method.
  The proposed solution finds a globally optimal point without the aid of the alternating optimization and the multiple initial points, and thus is efficient.
  The solution also can be exploited as a useful outerbound for the MuMIMO-CR-SWIPT system by regarding the multiple S-ID users as a single macro user
  with ideal multiuser cooperation.

  \item In Section \ref{sec:Discussion}, we provide an in-depth discussion on the proposed designs from practical implementation perspectives such as
  the required CSI at each node, the channel estimation procedure, and the computational complexity.

  \item Finally, in Section \ref{sec:Numerical Results}, we offer extensive simulation results to demonstrate the efficiency of the proposed designs.
  We first confirm that the proposed SuMIMO design attains the global optimum.
  Then, we verify the optimality of the MuMIMO design by observing
  that the performance approaches its SuMIMO outerbound with the aid of multiple initial points.
  One interesting observation is that the optimal point is achievable with only a few initial points in the low-to-medium SNR region,
  although a larger number of initial points may be still needed as SNR grows high.
  Obviously, the proposed MuMIMO design based on the WSR utility achieves the best WSR performance.
  However, the PF and HMR designs may be preferred over the WSR design in terms of the rate balancing among the S-ID users.
\end{itemize}


{\it Notations:} Throughout the paper, boldface upper and lowercase letters denote matrices and vectors, respectively.
The superscripts $(\cdot)^{\transp}$ and $(\cdot)^{\mathsf{H}}$ stand for the transpose and Hermitian-transpose operations, respectively.
We use $\mathsf{E}[\cdot]$, $\det(\cdot)$, and $\trace(\cdot)$ to denote the expectation, determinant, and trace operations, respectively.
The notation $\text{blkdiag}\{\Av_1,\ldots,\Av_K\}$ represents a blockwise diagonal matrix with matrices $\Av_1,\ldots,\Av_K$.
For a matrix $\Av$, we define $\delta_{\max}(\Av)$, $\delta_{\min}(\Av)$, $(\Av)_+$, and $\nabla f(\Av)$ as
the largest eigenvalue, the smallest eigenvalue, the element-wise $\max(\cdot,0)$ operation, and the gradient of $f(\cdot)$ at $\Av$, respectively.
Also, we define $\left[\{\Av_i\}_{i=1}^K\right]=[\Av_1,\ldots,\Av_K]$ as a matrix consisting of $\Av_i$'s from $i=1$ to $K$.
We define $\Iv_N$ as an $N\times N$ identity matrix.
Some important acronyms and notations are summarized in Table \ref{table:acronyms} and  \ref{table:definitions}, respectively.

\linespread{1.1}
\begin{table}
\centering \caption{Summary of Acronyms}
\vspace{-5pt}
\begin{tabular}{|c||c|}
\hline
Acronym   &  Definition   \\
\hline
\hline
CR   &  Cognitive Radio   \\
\hline
MMSE   &  Minimum Mean Squared Error   \\
\hline
WMMSE &  Weighted MMSE  \\
\hline
MIMO   &  Multi-input Multi-output  \\
\hline
MuMIMO &  Multiuser MIMO  \\
\hline
SuMIMO &  Single-user MIMO  \\
\hline
SNR &  Signal-to-Noise Ratio  \\
\hline
WSR &  Weighted Sum Rate  \\
\hline
PF &  Proportional Fairness  \\
\hline
HMR & Harmonic Mean Rate  \\
\hline
S-BS & Secondary Base Station  \\
\hline
S-ID & Secondary Information Decoding  \\
\hline
S-EH & Secondary Energy Harvesting  \\
\hline
P-ID & Primary Information Decoding  \\
\hline
CSI & Channel State Information  \\
\hline
\end{tabular}
\label{table:acronyms}
\vspace{-10pt}
\end{table}
\linespread{1.0}

\section{System Model}\label{sec:system model}

As shown in Fig. \ref{figure:SystemModel.eps}, we consider a general MuMIMO-CR-SWIPT network
where an unlicensed S-BS with $M$ antennas supports $K_I$ S-ID users and $K_E$ S-EH users by
sharing the licensed spectrum assigned to the $K_P$ P-ID users
such that the performance degradation of each active primary link is within a tolerable margin.
It is generally assumed that each of the S-ID, S-EH, and P-ID users has $N_I$, $N_E$, and $N_P$ number of antennas, respectively.
Here, we assume that $M$ is sufficiently large such that $M>K_PN_P$ to circumvent a feasibility issue
for the zero-interference conditions as will be described in more detail in Section \ref{sec:zero interference design}.

Define $\xv=[\xv_1^\transp,\ldots,\xv_{K_I}^\transp]^\transp\sim \mathcal{CN}(0, \Iv_{K_IN_I})$ and $\nv=[\nv_1^\transp,\ldots,\nv_{K_I}^\transp]^\transp\sim \mathcal{CN}(0, \sigma^2_n \Iv_{K_IN_I})$
as the baseband signal vectors for the data and noise associated with the S-ID users, respectively.
Then, considering the narrow-band flat fading channels,
the received signal vector $\yv=[\yv_1^\transp,\ldots,\yv_{K_I}^\transp]^\transp\in \mathbb{C}^{K_IN_{I}}$
for $K_I$ S-ID users can be expressed as
\bea\label{eq:Signal Model2}
\yv=\Hv\Fv\xv+\nv
\eea
where $\Hv\in\mathbb{C}^{K_IN_I\times M}$ and $\Fv\in\mathbb{C}^{M\times K_IN_I}$ denote the channel and the precoding matrices
from the S-BS to the S-ID users, respectively. Specifically, we have
\bea
\Hv=\left[\{\Hv_k^\transp\}_{k=1}^{K_I}\right]^\transp
~~\text{and}~~\Fv=\left[\{\Fv_k\}_{k=1}^{K_I}\right]\nonumber
\eea
where $\Hv_{k}\in\mathbb{C}^{N_{I}\times M}$ and $\Fv_k\in \mathbb{C}^{M\times N_{I}}$ represent
the channel and precoding matrices from the S-BS to the $k$-th S-ID user, respectively.
Thus, the received signal at the $k$-th S-ID $\yv_k$ can be rephrased by
\bea
\yv_k=\Hv_k\Fv_k\xv_k+\sum_{m=1,m\ne k}^{K_I} \Hv_k\Fv_m\xv_m + \nv_k,\nonumber
\eea
which leads to the information rate $R_k$ to the $k$-th S-ID user as
\bea\label{eq:rate}
R_k=\log\det(\Fv_k^\mathsf{H}\Hv_k^\mathsf{H}\Rv_{n,k}^{-1}\Hv_k\Fv_k+\Iv_{N_I})
\eea
where $\Rv_{n,k}\triangleq\sum_{m\ne k} \Hv_k\Fv_m\Fv_m^\mathsf{H}\Hv_k^\mathsf{H}+\sigma_{n}^{2}\Iv_{N_I}$ denotes the effective noise covariance matrix.
For simplicity, here we ignored the interference from the primary transmitter to the S-ID users, but the result can be applied to more general cases.
For notational convenience, we also define a stacked noise covariance as $\Rv_n\triangleq\text{blkdiag}[\Rv_{n,1},\ldots,\Rv_{n,K_I}]$.

\linespread{1.3}
\begin{table}
\centering \caption{Summary of Notations}
\vspace{-5pt}
\begin{tabular}{|c||c|}
\hline
Notation   &  Definition  \\
\hline
\hline
$M$   &  Number of S-BS antennas \\
\hline
$N_I$   &  Number of antennas at each S-ID user \\
\hline
$N_E$   &  Number of antennas at each S-EH user \\
\hline
$N_P$   &  Number of antennas at each P-ID user \\
\hline
$K_I$   &  Number of S-ID users \\
\hline
$K_E$   &  Number of S-EH users \\
\hline
$K_P$   &  Number of P-ID users \\
\hline
$E_{\text{th},i}$   &  Energy threshold at $i$-th S-EH user \\
\hline
$I_{\text{th},j}$   &  Interference threshold at $j$-th P-ID user \\
\hline
$\Hv_k$   &  Channel between S-BS and $k$-th S-ID user \\
\hline
$\Gv_i$   &  Channel between S-BS and $i$-th S-EH user \\
\hline
$\Tv_j$   &  Channel between S-BS and $j$-th P-ID user \\
\hline
$\Rv_{n,k}$   &  Effective noise covariance matrix at $k$-th S-ID user \\
\hline
$\Fv_k$   &  Precoding matrix for $k$-th S-ID user \\
\hline
$\Lv_k$   &  Receiver matrix for $k$-th S-ID user \\
\hline
$\Wv_k$   &  Weight matrix for $k$-th S-ID user \\
\hline
$\Cv_k$   &  MSE matrix for $k$-th S-ID user \\
\hline
$\Hv$   &  $[\Hv_1^{\transp},\ldots,\Hv_{K_I}^{\transp}]^{\transp}$\\
\hline
$\Tv$   &  $[\Tv_1^{\transp},\ldots,\Tv_{K_P}^{\transp}]^{\transp}$\\
\hline
$\Rv_n$   &  $\text{blkdiag}[\Rv_{n,1},\ldots,\Rv_{n,K_I}]$   \\
\hline
$\Fv$   &  $[\Fv_1,\ldots,\Fv_{K_I}]$\\
\hline
$\Lv$   &  $\text{blkdiag}\{\Lv_{1},\ldots,\Lv_{K_I}\}$   \\
\hline
$\Wv$   &  $\text{blkdiag}\{\Wv_{1},\ldots,\Wv_{K_I}\}$   \\
\hline
\end{tabular}
\label{table:definitions}
\vspace{-10pt}
\end{table}
\linespread{1.0}

Define the downlink channel matrices from the S-BS to the $i$-th S-EH and the $j$-th P-ID users as $\Gv_i\in \mathbb{C}^{N_E\times M}$ and
$\Tv_j\in \mathbb{C}^{N_P\times M}$, respectively.
Then, by employing the conventional linear EH model\footnote{Note that the proposed algorithm also works for a practical non-linear EH model as we consider individual EH constraints. Please refer to \cite{EBoshkovska:15} for more details.} in \cite{RZhang:13}, the amount of energy that can be harvested per a unit time at the $i$-th EH-user is quantified as
$\rho\Vert\Gv_i\Fv\Vert_{F}^2=\rho\trace(\Fv^\mathsf{H}\Gv_i^\mathsf{H}\Gv_i\Fv)$ where $0<\rho<1$ represents the RF-to-energy conversion efficiency.
For ease of presentation, we set $\rho=1$ unless stated otherwise.
Similarly, one can define the total interference power at the $j$-th P-ID user as $\Vert\Tv_j\Fv\Vert^2_{F}=\trace(\Fv^\mathsf{H}\Tv_j^\mathsf{H}\Tv_j\Fv)$ \cite{KyoungJae:11b}.

We consider the quasi-static fading environment, where the channel matrices are approximately constant over a few transmission blocks.
Then, considering the time division duplex (TDD) scheme,
the S-BS obtains the channel state information (CSI) of all links utilizing the uplink reference signals from the users, while each S-ID user obtains its own CSI by leveraging the downlink training from the S-BS.
Then, we can formulate a precoder design problem for sum-utility maximization in the MuMIMO-CR-SWIPT networks as
\bea \label{eq:Original Problem}
\text{(P-1)}~~\max_{{\bf{F}}} \sum_{k=1}^{K_I} U_k(R_k)~~~~~~~~~~~~~~~~~~~~~~~~~~~~~~~~~~~~~~\nonumber\\
\mbox{s.t.}~C_{\text{BS}}: \trace\big( \Fv^{\mathsf{H}}\Fv \big) \leq  P_T,\nonumber ~~~~~~~~~~~~~~~~~~~~~~~~~~~~~~~~~\\
C_{\text{EH}}:\trace\big(\Fv^\mathsf{H}\Gv_i^\mathsf{H}\Gv_i\Fv\big) \ge E_{\text{th},i} ~\text{for}~i=1,\ldots,K_E~~~\nonumber\\
C_{\text{CR}}:\trace\big(\Fv^\mathsf{H}\Tv_j^\mathsf{H}\Tv_j\Fv\big) \le I_{\text{th},j} ~\text{for}~j=1,\ldots,K_P~~~\nonumber
\eea
where $C_{\text{BS}}$ denote the transmit power constraint at the S-BS, and $C_{\text{EH}}$ and $C_{\text{CR}}$ represent
individual harvested energy and interference constraints for the S-EH and P-ID users, respectively.
Here, $E_{\text{th},i}$ and $I_{\text{th},j}$ refer to the target energy level at the $i$-th S-EH user and the target interference level at the $j$-th P-ID user, respectively.
$U_k(\cdot)$ indicates a utility function that is for example given by
$U_k(R_k)=\alpha_kR_k$, $U_k(R_k)=\log R_k$, and $U_k(R_k)=-R_k^{-1}$ for the WSR, PF, and HMR, respectively.
Note that (P-1) is generally non-convex, and thus is difficult to solve in its current form.
Throughout the paper, we assume that the S-BS solves (P-1) with global perfect CSIs of $\{\Hv_k,\Gv_i,\Tv_j,\forall k,i,j\}$.
More details about the required CSIs at each node and the corresponding channel acquisition procedure
will be discussed later in Section \ref{sec:Discussion}.

\section{Achievable Energy Region}\label{sec:Maximum Achievable Energy}

When the EH requirements at the S-EH users grow too high, the system may become infeasible due to
limited transmit power at the S-BS.
Therefore, it is important to check whether the system is feasible or not before solving the problem in (P-1).
In this section, we formulate an weighted sum harvested energy maximization problem in the CR-SWIPT topology to identify
the Pareto optimal boundary points of the achievable energy region in (P-1) and provide an efficient algorithm to find a solution.

Let us set $w_i\geq0$ as an weight factor for the harvested energy at the $i$-th EH user such that $\sum_{i=1}^{K_E}w_i=1$.
Then, the weighted sum harvested energy maximization problem can be formulated as
\bea\label{eq:max-energy problem}
\text{(P-2)}&&\min_{\mathbf{S}\succeq\mathbf{0}}~-\sum_{i=1}^{K_E}w_i\trace\big(\Gv_i\Sv\Gv_i^\mathsf{H}\big)\nonumber\\
\mbox{s.t.}&&\trace\big( \Sv \big) \leq  P_T,\nonumber ~~~~~~~~~~~~~~~~~~~~~~~~~~~~~~~~~\\
&&\trace\big(\Tv_j\Sv\Tv_j^\mathsf{H}\big) \leq I_{\text{th},j},~\forall j\nonumber.
\eea
The above problem is convex, for which the strong duality holds.
Here, we have relaxed a constraint $\Sv=\Fv\Fv^\mathsf{H}$, but one may recognize from the following proposition
that a precoding matrix that is given in the form $\Fv=[\fv~\mathbf{0}_{M\times (K_IN_I-1)}]$ can achieve the optimal value of (P-2).
According to the weight factors $w_i$'s, the resulting solution identifies each Pareto optimal boundary point of the achievable energy region.

\begin{Proposition}\label{Prop:Prop1}
With the assumption that $M>K_PN_P$, the optimum in (P-2) can be achieved by a rank-1 matrix $\Sv$ having
full transmit power $\trace(\Sv)=P_T$.
\end{Proposition}
\begin{IEEEproof}
First, by contradiction, let us presume that the optimum of (P-2) occurs at a point where $\trace(\Sv)<P_T$.
Define an aggregated P-ID user channels $\Tv=[\{\Tv_j^\transp\}_{j=1}^{K_P}]^\transp$.
Then, for $M>K_PN_P$, we can find a matrix $\Qv\in\mathbb{C}^{M\times(M-K_PN_P)}$ to meet $\Tv\Qv=\mathbf{0}$.
Therefore, any matrix $\Sv^\prime$ in the form of $\Sv^\prime=\Sv+c\Qv\Qv^\mathsf{H}$ with a constant $c>0$ such that $\trace(\Sv^\prime)=P_T$
achieves a greater amount of energy than $\Sv$ without violating all the constraints in (P-2),
which contradicts to our previous presumption. Therefore, $\trace(\Sv)=P_T$ is always optimal.

In the meantime, let us consider the Lagrangian as
\bea\label{eq:Lagrangian P2}
\mathcal{L}_{P2}=-\sum_{i=1}^{K_E}w_i\trace(\Gv_i\Sv\Gv_i^\mathsf{H})+\nu(\trace(\Sv)-P_T)-\trace(\mathbf{\Psi}\Sv)\nonumber\\
+\sum_{j=1}^{K_P}\mu_j(\trace(\Tv_j\Sv\Tv_j^\mathsf{H})-I_{\text{th},j})~~~~~~~~
\eea
where $\nu\ge0$, $\mu_j\ge0$, and $\mathbf{\Psi}\in\mathbb{C}^{M\times M}\succeq\mathbf{0}$ denote the dual variables
corresponding to the S-BS power constraint, the $j$-th CR constraint, and the semi-definite constraint, i.e., $\Sv\succeq\mathbf{0}$, respectively.

Then, from the KKT conditions, we have
\bea\label{eq:app1 optimality condition}
\mathbf{\Psi}=\nu\Iv_M-\Pv\succeq\mathbf{0}~~\text{and}~~\mathbf{\Psi}\Sv=\mathbf{0},
\eea
where $\Pv\triangleq \sum_{i=1}^{K_E}w_i\Gv_i^\mathsf{H}\Gv_i-\sum_j\mu_j\Tv_j^\mathsf{H}\Tv_j$.
From the former condition in (\ref{eq:app1 optimality condition}), we have $\nu\geq\delta_{\max}(\Pv)$. However, the latter condition, i.e., $\mathbf{\Psi}\Sv=\mathbf{0}$, only holds for $\nu=\delta_{\max}(\Pv)$ because otherwise $\mathbf{\Psi}$ becomes a full-rank matrix for which there exists no $\Sv\neq\mathbf{0}$ that satisfies $\mathbf{\Psi}\Sv=\mathbf{0}$.
Therefore, we can conclude that $\nu=\delta_{\max}(\Pv)$ is optimal,
which implies that the optimal $\Sv$ occurs at a point where all column vectors of $\Sv$
are aligned with the eigenvector corresponding to $\delta_{\max}(\Pv)$, and the proof is concluded.
\end{IEEEproof}

The result in Proposition 1 enables us to find a solution of (P-2) without solving the complicated SDP problem.
Specifically, let us define $\uv=\big[\{\mu_j\}_{j=1}^{K_P}\big]$.
Then, by leveraging (\ref{eq:Lagrangian P2}) and (\ref{eq:app1 optimality condition}),
we can formulate a simple dual problem of (P-2) as
\bea
\text{(P-3)}~~~~\sup_{\mathbf{u}\succeq\mathbf{0}}~~g(\uv)~~~\nonumber
\eea
where $g(\uv)\triangleq\inf_{\mathbf{S}\succeq\mathbf{0}}\mathcal{L}_{P2}=-\delta_{\max}(\Pv)P_T-\sum_{j=1}^{K_P}\mu_jI_{\text{th},j}$,
which is easily solved via the subgradient ellipsoid method \cite{SBoyd:18},
for which it can be shown that the subgradient of $g(\uv)$ at a point $\uv$ is given by
$\{-\Vert\Tv_j\pv\Vert^2+I_{\text{th},j}\}_{j=1}^{K_P}$ \cite{Golub:96}.
Here, $\pv$ denotes the eigenvector of $\Pv$ corresponding to $\delta_{\max}(\Pv)$.
After we find the dual optimal $\uv$, the primary optimal precoding matrix $\Fv=[\fv~\mathbf{0}_{M\times (K_IN_I-1)}]$ is computed
such that $(\delta_{\max}(\Pv)\Iv_M-\Pv)\fv=\mathbf{0}$ with $\Vert\fv\Vert^2=P_T$ from Proposition 1.
The algorithm is summarized in Table \ref{table:alg for (P-3)}.

\begin{table}
\centering \caption{Algorithm for solving (P-3)}
\begin{tabular}{ | l |}
\hline
{Initialize} $\uv\succeq\mathbf{0}$. \\
{\bf Repeat} \\
{~~~Compute} the subgradient of $g(\uv)$ for a given $\uv$.\\
{~~~Update} $\uv$ using the ellipsoid method subject to $\uv\succeq\mathbf{0}$.\\
{\bf Until} $\uv$ converges to the prescribed accuracy.\\
{Find} $\fv$ such that  $(\delta_{\max}(\Pv)\Iv_M-\Pv)\fv=\mathbf{0}$ and $\Vert\fv\Vert^2=P_T$.\\
\hline
\end{tabular}
\label{table:alg for (P-3)}
\end{table}

\section{Proposed MuMIMO-CR-SWIPT Precoder Designs}\label{sec:Sum-Utility Maximization}
In this section, we provide an efficient optimization algorithm to solve the sum-utility maximization problem in (P-1) for
general MuMIMO-CR-SWIPT networks.

\subsection{WMMSE Problem Reformulation}

First, we transform (P-1) to an equivalent WMMSE problem that is relatively easy to solve.
Let us define $\Lv_k \in \mathbb{C}^{N_{I}\times N_{I}}$ and $\hat{\yv}_k=\Lv_k\yv_k$
as the linear receiver and the final observation at the $k$-th S-ID user, respectively.
Then, one can compute the MSE matrix as
\bea \label{eq:MSE matrix}
\Cv_{k} \triangleq   \mathsf{E}[(\gamma^{-1}\hat{\yv}_k-\xv_k)(\gamma^{-1}\hat{\yv}_k-\xv_k)^H]~~~~~~~~~~\nonumber \\
=\gamma^{-2}\Lv_k \big(\Hv_k\Fv_k\Fv_k^\mathsf{H}\Hv_k^\mathsf{H}+\Rv_{n,k}\big)\Lv_k^H~~~~~\nonumber\\
-\gamma^{-1}\Lv_k\Hv_k\Fv_k - \gamma^{-1}\Fv_k^\mathsf{H}\Hv_k^\mathsf{H}\Lv_k^\mathsf{H} + \Iv_{N_{I}},
\eea
where a new variable $\gamma>0$ enables us to obtain an efficient algorithm.

For convenience, let us set $\Fv=\gamma\bar{\Fv}$ for unknown $\bar{\Fv}$.
Then, by introducing an weight matrix $\Wv_k\in \mathbb{C}^{N_{I}\times N_{I}}$,
we can reformulate (P-1) to an equivalent WMMSE problem as
\bea \label{eq:WMMSE problem}
(\text{P-4})~~\inf_{\gamma,\bar{\mathbf{F}},\{\mathbf{W}_k,\mathbf{L}_k,\forall k\}}\sum_{k=1}^{K_I}\big\{\trace(\Wv_k\Cv_{k}) + e_k(\Wv_k)\big\} \nonumber\\
\mbox{s.t.}~\bar{C}_{\text{BS}}: \trace\big( \bar{\Fv}^{\mathsf{H}}\bar{\Fv} \big) \leq  \gamma^{-2}P_T,\nonumber ~~~~~~~~~~~~~~~~~~~~~\\
\bar{C}_{\text{EH}}:\trace\big(\bar{\Fv}^\mathsf{H}\Gv_i^\mathsf{H}\Gv_i\bar{\Fv}\big) \ge \gamma^{-2}E_{\text{th},i}, ~\forall i~~~~~~~~~\nonumber\\
\bar{C}_{\text{CR}}:\trace\big(\bar{\Fv}^\mathsf{H}\Tv_j^\mathsf{H}\Tv_j\bar{\Fv}\big) \le \gamma^{-2}I_{\text{th},j}, ~\forall j~~~~~~~~~\nonumber
\eea
where $e_k(\Wv_k)\triangleq \eta_k\big(\mathbf{\Gamma}_k(\Wv_k)\big)-\text{Tr}\big(\Wv_k^\transp\mathbf{\Gamma}_k(\Wv_k)\big)$ and $\eta_k(\cdot)\triangleq-U_k(-\log\det(\cdot))$.
Here, $\mathbf{\Gamma}_k(\cdot)$ denotes the inverse mapping of the gradient map $\nabla\eta_k(\cdot)$,
e.g., $\nabla\eta_k\big(\mathbf{\Gamma}_k(\Wv_k)\big)=\mathbf{\Gamma}_k(\nabla\eta_k(\Wv_k))=\Wv_k$.
As long as $\eta_k(\cdot)$ is a strictly concave function for all $k$, the equivalence between (P-1) and (P-4) holds.
Detailed proof simply follows from \cite[Section II-B]{QShi:11}.

Although (P-4) is still jointly non-convex,
it is now seen as an unconstrained convex problem with respect to each of $\Wv\triangleq\text{blkdiag}\{\Wv_1,\ldots,\Wv_{K_I}\}$
and $\Lv\triangleq\text{blkdiag}\{\Lv_1,\ldots,\Lv_{K_I}\}$ for given $\gamma$ and $\bar{\Fv}$.
Therefore, the optimal structures of $\Wv_k$ and $\Lv_k$ are easily acquired from the KKT zero-gradient conditions.

Let us set the Lagrangian of (P-4) as
\bea \label{eq:Lagrangian P4}
\mathcal{L}_{P4}=\trace(\Wv\Cv)+e(\Wv)+\nu\big(\trace(\bar{\Fv}\bar{\Fv}^\mathsf{H})-\gamma^{-2}P_T\big)\nonumber\\
-\sum_{i=1}^{K_E}\lambda_i\big(\trace(\bar{\Fv}^\mathsf{H}\Gv_i^\mathsf{H}\Gv_i\bar{\Fv})-\gamma^{-2}E_{\text{th},i}\big)\nonumber\\
+\sum_{j=1}^{K_P}\mu_j\big(\trace(\bar{\Fv}^\mathsf{H}\Tv_j^\mathsf{H}\Tv_j\bar{\Fv})-\gamma^{-2}I_{\text{th},j}\big)
\eea
where $\nu\ge0$, $\lambda_i\ge0$, and $\mu_j\ge0$ denote the dual variables
corresponding to $\bar{C}_{\text{BS}}$, the $i$-th $\bar{C}_{\text{EH}}$, and $j$-th $\bar{C}_{\text{CR}}$ constraints, respectively.
Also, here we define $e(\Wv)\triangleq\sum_{k=1}^{K_I}e_k(\Wv_k)$ and
\bea
\Cv&\triangleq&\text{blkdiag}\{\Cv_{1},\ldots,\Cv_{K_I}\}\nonumber\\
&=&\Lv\big(\Hv\bar{\Fv}\bar{\Fv}^\mathsf{H}\Hv^\mathsf{H}+\gamma^{-2}\sigma_{n}^{2}\Iv_{K_IN_I}\big)\Lv^\mathsf{H}\nonumber\\
&&-\Lv\Hv\bar{\Fv} - \bar{\Fv}^\mathsf{H}\Hv^\mathsf{H}\Lv^\mathsf{H} + \Iv_{K_IN_I}.
\eea

Then, the KKT necessary conditions for optimality are given by
\bea
\label{eq:KKT1}
\Lv_k \big(  \Hv_k\bar{\Fv}_k\bar{\Fv}_k^\mathsf{H}\Hv_k^\mathsf{H}+\gamma^{-2}\Rv_{n,k} \big)=\bar{\Fv}_k^\mathsf{H}\Hv_k^\mathsf{H},\forall k\\
\label{eq:KKT2}
\Cv_{k}^\transp-\mathbf{\Gamma}_k(\Wv_k)=\mathbf{0},\forall k\\
\label{eq:KKT3}
\big(\Hv^{\mathsf{H}}\Lv^\mathsf{H}\Wv\Lv\Hv-\sum_{i}\lambda_i\Gv_i^{\mathsf{H}}\Gv_i+\sum_{j}\mu_j\Tv_j^{\mathsf{H}}\Tv_j+\nu\Iv_M\big)\bar{\Fv}\nonumber\\
=\Hv^{\mathsf{H}}\Lv^\mathsf{H}\Wv^\mathsf{H}\\
\label{eq:KKT4}
\beta+\sum_{i}\lambda_iE_{\text{th},i}-\sum_{j}\mu_jI_{\text{th},i}=\nu P_T\\
\label{eq:KKT constraint}
\bar{C}_{\text{BS}};~\bar{C}_{\text{EH}};~\bar{C}_{\text{CR}}\\
\label{eq:Slack BS}
\nu\big(\trace(\bar{\Fv}^{\mathsf{H}}\bar{\Fv})-\gamma^{-2}P_T\big)=0\\
\label{eq:Slack EH}
\lambda_i\big(\trace(\bar{\Fv}^\mathsf{H}\Gv_i^\mathsf{H}\Gv_i\bar{\Fv})-\gamma^{-2}E_{\text{th},i}\big)=0,\forall i\\
\label{eq:Slack CR}
\mu_j\big(\text{Tr}(\bar{\Fv}^{\mathsf{H}}\Tv_j^\mathsf{H}\Tv_j\bar{\Fv})-\gamma^{-2}I_{\text{th},j}\big)=0,\forall j
\eea
where $\bar{\Fv}_k=\gamma^{-1}\Fv_k$ and $\beta=\text{Tr}(\sigma_n^2\Wv\Lv\Lv^\mathsf{H})$.
Here, the equations from (\ref{eq:KKT1}) to (\ref{eq:KKT4}) stem from the zero gradient conditions with respect to $\Lv_k$, $\Wv_k$, $\bar{\Fv}$, and $\gamma$, respectively,
and the equations from (\ref{eq:Slack BS}) to (\ref{eq:Slack CR}) represent the complement slackness conditions.
Note that (\ref{eq:KKT2}) follows from \cite[Theorem 2]{QShi:11}.

By condition (\ref{eq:KKT1}), we find the optimal receiver $\Lv_{k}$ as
\bea \label{eq:WMMSE_Receiver}
\Lv_{k}=\bar{\Fv}_k^\mathsf{H}\Hv_k^\mathsf{H}\big(\Hv_k\bar{\Fv}_k\bar{\Fv}_k^\mathsf{H}\Hv_k^\mathsf{H}+\gamma^{-2}\Rv_{n,k} \big)^{-1},\forall k
\eea
which in turn makes the MSE matrix in (\ref{eq:MSE matrix}) given in a compact form of
$\Cv_{L,k}=(\gamma^2\bar{\Fv}_k^\mathsf{H}\Hv_k^\mathsf{H}\Rv_{n,k}^{-1}\Hv_k\bar{\Fv}_k+\Iv)^{-1}$ \cite{Palomar:03}.
Then, from (\ref{eq:KKT2}), we can update the optimal weight matrix as
\bea \label{eq:Weight matrix}
\Wv_k=\nabla\eta_k(\Cv_{L,k}^\transp), \forall k.
\eea
For instance, we have $\Wv_k=\alpha_k\Cv_{L,k}^{-1}$, $\Wv_k=\big(-\big(\log\det(\mathbf{C}_{L,k})\big)\Cv_{L,k}\big)^{-1}$, and $\Wv_k=\big(\big(\log\det(\mathbf{C}_{L,k})\big)^2\Cv_{L,k}\big)^{-1}$ according to our target utilities WSR, HMR, and PF in (P-1), respectively.

\subsection{Optimal Precoder Design}\label{sec:Max-rate Design}

Unlike the conventional non-SWIPT designs, (P-4) is still non-convex with respect to $\gamma$ and $\bar{\Fv}$,
because the conflicting constraints in $\bar{C}_{\text{BS}}$, $\bar{C}_{\text{EH}}$, and $\bar{C}_{\text{CR}}$
form a non-convex feasible domain. Thus, standard CVX tools
such as SeDuMi \cite{SBoyd:04} are not immediately applicable even if other variables $\Wv$ and $\Lv$ are fixed.
Therefore, it is most important to determine the optimal structure of $\gamma$ and $\bar{\Fv}$.
Once their optimal forms are identified, (P-4) is easily solved
by alternately updating $\gamma$, $\bar{\Fv}$, and $\{\Wv_k,\Lv_k,\forall k\}$ until convergence.

First, we observe from (\ref{eq:KKT4}) that for a fixed $\beta$, $\nu$ in (\ref{eq:Lagrangian P4}) is expressed as a function of $\bar{\uv}\triangleq[\lambda_1,\ldots,\lambda_{K_E},\mu_1,\ldots,\mu_{K_P}]$.
Thus, we can reduce the number of unknown dual variables by $1$.
Also, as we have $\beta>0$, at least one variable among $\{\nu,\mu_1,\ldots,\mu_{K_P}\}$ has a non-zero positive value,
which implies that at least one constraint in $\bar{C}_{\text{BS}}$ and $\bar{C}_{\text{CR}}$ must be activated due to (\ref{eq:Slack BS}) and (\ref{eq:Slack CR}).
Thus, for a given $\bar{\Fv}$, we have
\bea\label{eq:gamma}
\gamma=\sqrt{\min\left(\frac{P_T}{\trace(\bar{\mathbf{F}}^\mathsf{H}\bar{\mathbf{F}})},
\left\{\frac{I_{\text{th},j}}{\trace(\bar{\mathbf{F}}^\mathsf{H}\mathbf{T}_j^\mathsf{H}\mathbf{T}_j\bar{\mathbf{F}})}\right\}_{j=1}^{K_P}\right)}.
\eea

Next, we consider a Lagrange dual function of (P-4) as
\bea\label{eq:dual function P4}
h(\bar{\uv})=\inf_{\gamma,\bar{\mathbf{F}}}\mathcal{L}_{P4}(\nu,\bar{\uv},\gamma,\bar{\Fv})
=\inf_{\bar{\mathbf{F}}}\bar{\mathcal{L}}_{P4}(\bar{\uv},\bar{\Fv}),
\eea
where $\bar{\mathcal{L}}_{P4}(\bar{\uv},\bar{\Fv})$ is obtained by applying (\ref{eq:KKT4}) to (\ref{eq:Lagrangian P4}) as
\bea\label{eq:new Lagrangian}
\bar{\mathcal{L}}_{P4}=\trace\big(\Wv\Lv\Hv\bar{\Fv}\bar{\Fv}^\mathsf{H}\Hv^\mathsf{H}\Lv^\mathsf{H}
-\Wv\Lv\Hv\bar{\Fv}~~~~~~~~~~~~~~ \nonumber\\
- \Wv\bar{\Fv}^\mathsf{H}\Hv^\mathsf{H}\Lv^\mathsf{H}+\Wv\big)+ e(\Wv)-\sum_{i}\lambda_i\trace(\bar{\Fv}^\mathsf{H}\Zv_{E,i}\bar{\Fv})\nonumber\\
+\sum_{j}\mu_j\trace(\bar{\Fv}^\mathsf{H}\Zv_{P,j}\bar{\Fv})+\frac{\beta\trace(\bar{\Fv}^\mathsf{H}\bar{\Fv})}{P_T}.~~~~~~~~
\eea
Here, we define $\Zv_{E,i}\triangleq\Gv_i^\mathsf{H}\Gv_i-\frac{E_{\text{th},i}}{P_T}\Iv_M$ and $\Zv_{P,j}\triangleq\Tv_j^\mathsf{H}\Tv_j-\frac{I_{\text{th},j}}{P_T}\Iv_M$.
Now, let us temporarily ignore the constant terms in $\bar{\mathcal{L}}_{P4}$ with respect to $\bar{\Fv}$, which makes (\ref{eq:dual function P4}) rephrased by
\bea\label{eq:dual function}
\inf_{\bar{\mathbf{F}}}~\Big\{\text{Tr}\big(\bar{\mathbf{F}}^{\mathsf{H}}\Kv\bar{\mathbf{F}}\big)-
\text{Tr}\big(\Wv\Lv\Hv\bar{\mathbf{F}}\big)\Big\},
\eea
where $\Kv\triangleq\Yv-\sum_i\lambda_i\Zv_{E,i}+\sum_j\mu_j\Zv_{P,j}$ with $\Yv\triangleq\Hv^\mathsf{H}\Lv^\mathsf{H}\Wv\Lv\Hv+\frac{\beta}{P_T}\Iv_M$.
Now, suppose that at least one eigenvalue of $\Kv$ is non-positive with corresponding eigenvector $\vv\in\mathbb{C}^{M\times1}$.
Then, we can make (\ref{eq:dual function}) unbounded from below by simply setting $\mathbf{F}=[\fv ~\mathbf{0}_{M\times(K_IN_I-1)}]$
with $\Vert\fv\Vert^2=\infty$.
Therefore, a dual feasible condition $\Kv\succ\mathbf{0}$ arises for (\ref{eq:dual function P4}),
which leads us to the following dual problem as
\bea\label{eq:dual problem}
\text{(P-5)}~~~~~\sup_{\bar{\mathbf{u}}\succeq\mathbf{0}}~~ h(\bar{\mathbf{u}})~~\mbox{s.t.}~~\delta_{\min}(\Kv)>0.\nonumber
\eea

\begin{Proposition}\label{Prop:Prop2}
There exists zero-duality gap between (P-4) and its dual (P-5)
with respect to $\gamma$ and $\bar{\Fv}$.
\end{Proposition}
\begin{IEEEproof}
See Appendix \ref{Appendix A}.
\end{IEEEproof}

With the assistance of Proposition \ref{Prop:Prop2}, we can find optimal $\bar{\Fv}$ through (P-5) that is solvable via the ellipsoid method for constrained problems \cite{SBoyd:18}, for which
the subgradient of $h(\bar{\uv})$ at a feasible point $\bar{\uv}$ is computed by $[\{\trace(\bar{\Fv}^{\star\mathsf{H}}\Zv_{E,i}\bar{\Fv}^\star)\}_{i=1}^{K_E},\{-\trace(\bar{\Fv}^{\star\mathsf{H}}\Zv_{P,j}\bar{\Fv}^\star)\}_{j=1}^{K_P}]$.
Here, $\Fv^\star$ denotes the corresponding primal optimal solution.
Note that $\Kv$ is invertible for a feasible $\bar{\uv}$ since $\delta_{\min}(\Kv)>0$,
and thus we have
\bea\label{eq:Primal opt 0}
\bar{\Fv}^\star=\Kv^{-1}\Hv^\mathsf{H}\Lv^\mathsf{H}\Wv^\mathsf{H}.\nonumber
\eea
Otherwise if $\bar{\uv}$ violates the dual feasible condition, i.e., $\delta_{\min}(\Kv)\le0$, we compute the subgradient of $\delta_{\min}(\Kv)$ as
$[\{\kv^{\mathsf{H}}\Zv_{E,i}\kv\}_{i=1}^{K_E},\{-\kv^\mathsf{H}\Zv_{P,j}\kv\}_{j=1}^{K_P}]$ where $\kv\in\mathbb{C}^{M\times1}$ represents
the eigenvector of $\Kv$ corresponding to $\delta_{\min}(\Kv)$.
The algorithm is summarized in Table \ref{table:alg for (P-5)}.
After finding the optimal $\bar{\Fv}$, we finally set $\gamma$ as in (\ref{eq:gamma}),
which results in the optimal precoder $\Fv=\gamma\bar{\Fv}$.

\begin{table}
\centering \caption{Algorithm for solving (P-5)}
\begin{tabular}{ | l |}
\hline
{Initialize} $\bar{\uv}\succeq\mathbf{0}$. \\
{\bf Repeat} \\
{~~~Compute} $\Kv$ for a given $\bar{\uv}$.\\
{\bf~~~if} $\delta_{\min}(\Kv)>0$\\
{~~~~~~Compute} $\bar{\Fv}^\star=\Kv^{-1}\Hv^\mathsf{H}\Lv^\mathsf{H}\Wv^\mathsf{H}$.\\
{~~~~~~Compute} the subgradient of $h(\bar{\uv})$.\\
{\bf~~~else} \\
{~~~~~~Compute} the subgradient of $\delta_{\min}(\Kv)$.\\
{\bf~~~end} \\
{~~~Update} $\bar{\uv}$ using the ellipsoid method subject to $\bar{\uv}\succeq\mathbf{0}$.\\
{\bf Until} $\bar{\uv}$ converges to the prescribed accuracy.\\
{Set} $\bar{\Fv}=\bar{\Fv}^\star$.\\
\hline
\end{tabular}
\label{table:alg for (P-5)}
\end{table}

Algorithm \ref{algorithm:algorithm1} illustrates the entire WMMSE algorithm
for sum-utility maximization in general MuMIMO CR SWIPT systems with $K_I$ S-ID, $K_E$ S-EH, and $K_P$ P-ID users.
The algorithm converges, since each update of
$\Lv$, $\Wv$, and $\Fv$ minimizes the weighted sum-MSE that is bounded from below.
The converged point ensures the local optimum because all gradients with respect to $\Lv$, $\Wv$, and $\Fv$ simultaneously vanish.
Nevertheless, due to jointly non-convexity of (P-4), we may need $N_G$ different random initial points so that the resulting local minimum gets closer to the global minimum.
This may require additional outer-loop iterations.

\begin{algorithm}
\caption{Proposed MuMIMO CR SWIPT}
\begin{algorithmic}
\STATE Set target metric $U_k(\cdot)$.
\STATE Draw achievable energy region $(E_{\text{th},1},\ldots,E_{\text{th},K_E})$ from Table \ref{table:alg for (P-3)}.
\STATE Generate $N_G$ random initial points $\{\bar{\Fv}^{(1)},\ldots,\bar{\Fv}^{(N_G)}\}$.
\FOR{$i_p=1:N_G$}
    \STATE Initialize $\bar{\Fv}=\bar{\Fv}^{(i_p)}$ and compute $\gamma$ from (\ref{eq:gamma}).
    \REPEAT
        \STATE Compute $\Lv$ and $\Wv$ respectively from (\ref{eq:WMMSE_Receiver}) and (\ref{eq:Weight matrix}) for given $\gamma$ and $\bar{\Fv}$.
        \STATE Find $\bar{\Fv}$ from Table \ref{table:alg for (P-5)} for given $\Lv$ and $\Wv$.
        \STATE Compute $\gamma$ from (\ref{eq:gamma}) for a given $\bar{\Fv}$.
    \UNTIL{convergence.}
    \STATE Save $\Fv^{(i_p)}=\gamma\bar{\Fv}$.
\ENDFOR
\STATE Select the best one among $N_G$ different solutions $\{\Fv^{(i_p)}\}_{i_p=1}^{N_G}$.
\end{algorithmic}
\label{algorithm:algorithm1}
\end{algorithm}

\subsection{Zero-interference Design}\label{sec:zero interference design}

When $I_{\text{th},j}=0$ for some $j\in\mathbb{K}_P$ where $\mathbb{K}_P$ denotes a subset of P-ID user indices $\{1,2,\ldots,K_P\}$,
the algorithm in Section \ref{sec:Sum-Utility Maximization} may be inefficient and unstable,
because we may find unnecessary dual variables that are associated with the zero-interference constraints.
It is thus imperative to modify the optimization problem so that one can solve the problem more efficiently.

Define a stacked P-ID user channel matrix as
\bea\label{eq:stacked PU matrix}
\Tv_{\text{stack}}=[\{\Tv_j^\transp\}_{j\in\mathbb{N}}]^\transp\in\mathbb{C}^{QN_I\times M}
\eea
where $Q$ designates a cardinality of $\mathbb{N}$.
As we assume that $M>QN_I$, we can also define a matrix $\Uv\in\mathbb{C}^{M\times(M-QN_I)}$ whose column vectors constitute the orthonormal basis in the null-space of $\Tv_{\text{stack}}$, i.e., $\Tv_{\text{stack}}\Uv=\mathbf{0}$ with $\Uv^\mathsf{H}\Uv=\Iv_{M-QN_I}$.
Note that otherwise if $M\leq QN_I$, the system might be infeasible.

The precoding matrix that satisfies the zero-interference constraints,
i.e., $I_{\text{th},j}=0,\forall j\in\mathbb{N}$, must be in the null-space of $\Tv_{\text{stack}}$.
Therefore, without loss of optimality, the optimal precoder can be generally expressed by $\Fv=\Uv\tilde{\Fv}$
for any matrix $\tilde{\Fv}\in\mathbb{C}^{(M-QN_I)\times K_IN_I}$.
Thus, applying the result to (P-1), we obtain a modified optimization problem as
\bea
\text{(P-6)}~~\max_{{\tilde{\bf{F}}}} \sum_{k=1}^{K_I} U_k(\tilde{R}_k)~~~~~~~~~~~~~~~~~~~~~~~~~~\nonumber\\
\mbox{s.t.}~\trace\big( \tilde{\Fv}^{\mathsf{H}}\tilde{\Fv} \big) \leq  P_T,\nonumber ~~~~~~~~~~~~~~~~~~~~\\
\trace\big(\tilde{\Fv}^\mathsf{H}\tilde{\Gv}_i^\mathsf{H}\tilde{\Gv}_i\tilde{\Fv}\big) \ge E_{\text{th},i},\forall i~~~~~~~~\nonumber\\
\trace\big(\tilde{\Fv}^\mathsf{H}\tilde{\Tv}_j^\mathsf{H}\tilde{\Tv}_j\tilde{\Fv}\big) \le I_{\text{th},j},\forall j\in\mathbb{K}_P^\mathsf{C},\nonumber
\eea
where $\mathbb{K}_P^\mathsf{C}$ denotes a complementary set of $\mathbb{K}_P$ and $\tilde{R}_k=\log\det(\tilde{\Fv}_k^\mathsf{H}\tilde{\Hv}_k^\mathsf{H}\Rv_{n,k}^{-1}\tilde{\Hv}_k\tilde{\Fv}_k+\Iv_{N_I})$
with $\tilde{\Hv}\triangleq\Hv\Uv$, $\tilde{\Gv}\triangleq\Gv\Uv$, and $\tilde{\Tv}\triangleq\Tv\Uv$.
Once we find $\tilde{\Fv}$, the resulting solution $\Fv=\Uv\tilde{\Fv}$ achieves the zero interference constraints, i.e., $I_{\text{th},j}=0,\forall j\in\mathbb{K}_P^C$
with reduced number of dual variables by $Q$, and thus is efficient.
The rest of derivations is the same as the previous section.
Note that when we consider all zero interference, i.e., $I_{\text{th},j}=0, \forall j$,
the CR constraints in (P-6) is completely removed, which we call a {\it zero-forcing (ZF) design}.
Further, if we have $K_E=1$ and $U_k(\tilde{R}_k)=\tilde{R}_k$, (P-6) becomes equivalent to the one in \cite{CSong:16TWC}.

\section{Joint Optimal Solution for Single S-ID User} \label{sec:Joint optimal design}

In practice, the S-BS may support one S-ID user at a time in a TDMA manner.
In this case, our system model reduces to the SuMIMO channel with multiple CR and EH constraints,
for which we can show that the WMMSE problem in (P-4) can be jointly optimized without the aid of the multiple initial points
and the alternating optimization among the filters.
The result in this section not only provides the globally optimal solution for the SuMIMO-CR-SWIPT system,
but also serves as a theoretical performance outer bound for the MuMIMO-CR-SWIPT systems in the previous section.
Throughout the section, we will drop the S-ID user index $k$ from all variables related to the S-ID users,
since we only consider $K_I=1$.
Also, the auxiliary variable $\gamma$ in (\ref{eq:MSE matrix}) is now included in the receiver $\Lv$ from the joint optimization perspective.

\subsection{Joint Optimal Precoder Design}

Setting $\gamma=1$ and $\Fv=\bar{\Fv}$,
and plugging (\ref{eq:WMMSE_Receiver}) into (P-4),
we obtain a modified WMMSE problem for $K_I=1$ as
\bea \label{eq:WMMSE problem 2}
(\text{P-7})~~\inf_{\mathbf{F},\mathbf{W}}\trace\big(\Wv\Cv_L\big) + e(\Wv)~~~~~~~~\nonumber\\
\mbox{s.t.}~~~\trace\big( \Fv^{\mathsf{H}}\Fv \big) \leq  P_T,~~~~~~~~~~~~~\nonumber \\
\trace\big(\Fv^\mathsf{H}\Gv_i^\mathsf{H}\Gv_i\Fv\big) \ge E_{\text{th},i}, ~\forall i\nonumber\\
\trace\big(\Fv^\mathsf{H}\Tv_j^\mathsf{H}\Tv_j\Fv\big) \le I_{\text{th},j}, ~\forall j~\nonumber
\eea
where $\Rv_n=\sigma_n^2\Iv_{N_I}$, $\Cv_L=(\Fv^\mathsf{H}\Hv^\mathsf{H}\Rv_{n}^{-1}\Hv\Fv+\Iv_{N_I})^{-1}$, and
\bea\label{eq:e(Wv)}
e(\Wv)&=&\eta\big(\mathbf{\Gamma}(\Wv)\big)-\text{Tr}\big(\Wv^\transp\mathbf{\Gamma}(\Wv)\big)\nonumber\\
&=&-\log\det\big(\Wv\big)-N_I.
\eea
Here, (\ref{eq:e(Wv)}) follows, since for $K_I=1$ the general sum-utility maximization in (P-1) boils down to the rate maximization,
i.e., $\max_\mathbf{F}\sum_kU_k(R_k)\Rightarrow\max_\mathbf{F}\log\det(\Fv^\mathsf{H}\Hv^\mathsf{H}\Rv_{n}^{-1}\Hv\Fv+\Iv_{N_I})$,
for which the inverse mapping of $\nabla\eta(\cdot)$
is explicitly given by $\Gamma(\Wv)=\big(\Wv^\transp\big)^{-1}$.
Note that (P-7) is still jointly non-convex, and therefore the optimal solution is not immediate from there.

Define the Lagrangian for (P-7) as
\bea\label{eq:Lagrange P4-1}
\bar{\mathcal{L}}_{P4}=\trace\big(\Wv(\Fv^\mathsf{H}\Hv^\mathsf{H}\Rv_{n}^{-1}\Hv\Fv+\Iv_{N_I})^{-1}\big)+e(\Wv)\nonumber\\
+\trace\big(\Fv^\mathsf{H}\Mv\Fv\big)-\nu P_T+\sum_i\lambda_i E_{\text{th},i}-\sum_j\mu_j I_{\text{th},j}
\eea
with $\Mv\triangleq\nu\Iv_{M}-\sum_i\lambda_i\Gv_i^\mathsf{H}\Gv_i+\sum_j\mu_j\Tv_i^\mathsf{H}\Tv_i$.
We first see from (\ref{eq:Lagrange P4-1}) that the optimal weight matrix for a given $\Fv$ should be positive definite,
because we have
\bea\label{eq:Weight matrix 1}
\Wv&=&\nabla\eta(\Cv_L^\transp)\\
\label{eq:Weight matrix 2}
&=&\Fv^\mathsf{H}\Hv^\mathsf{H}\Rv_{n}^{-1}\Hv\Fv+\Iv_{N_I}
\eea
where (\ref{eq:Weight matrix 1}) follows from the optimality condition in (\ref{eq:Weight matrix}).
Based on the result, now we can find the optimal precoding structure $\Fv$ as described in the following proposition.

\begin{Proposition}\label{Prop:Prop3}
The optimal solution $\Fv$ in (P-7) for a given positive definite weight matrix $\Wv$ has the form of
\bea\label{eq:Primal opt}
\Fv^\star=\Mv^{-1/2}\Vv_1(\mathbf{W}^{1/2}\mathbf{\Phi}_1^{-1/2}-\mathbf{\Phi}_1^{-1})_+^{1/2},
\eea
where $\Vv_1\in\mathbb{C}^{M\times N_I}$ and $\mathbf{\Phi}_1\in\mathbb{C}^{N_I\times N_I}$ come from the following eigenvalue decomposition
\bea\label{eq:eigen-decomposition}
\Mv^{-\frac{1}{2}}\Hv^H\Rv_{n}^{-1}\Hv\Mv^{-\frac{1}{2}}=\Vv\mathbf{\Phi}\Vv^\mathsf{H}.
\eea
with a unitary matrix $\Vv=[\Vv_1~\Vv_2]\in\mathbb{C}^{M\times M}$ and a square diagonal matrix $\mathbf{\Phi}=\text{blkdiag}\{\mathbf{\Phi}_1~\mathbf{0}\}\in\mathbb{C}^{M\times M}$ having the eigenvalues of (\ref{eq:eigen-decomposition}) in a descending order.
\end{Proposition}
\begin{IEEEproof}
See Appendix \ref{Appendix B}.
\end{IEEEproof}

\subsection{Weight Matrix Design}

The weight matrix $\Wv$ can be designed differently according to the applications.

\subsubsection{Maximum Rate Design} \label{sec:Max-rate Design SuMIMO}
Due to the equivalence between (P-1) and (P-7),
the maximum rate can be achieved when both $\Fv$ and $\Wv$ jointly solve (P-7).
To this end, the optimal weight matrix must satisfy the equality in (\ref{eq:Weight matrix 2}) that is
alternatively expressed by using $\Fv$ in (\ref{eq:Primal opt}) as
\bea
\Wv-\Iv_{N_I}=(\Wv^{1/2}\mathbf{\Phi}_1^{1/2}-\Iv_{N_I})_+.
\eea
Thus, the $k$-th diagonal element of $\Wv$ should become
\bea
w_{k}=\left\{\begin{array}{cc}
                 \phi_{k} & \text{if}~~\phi_{k}\geq1\\
                 1 & \text{else}
               \end{array}\right.
\eea
where $\phi_{k}$ denotes the $k$-th diagonal element of $\mathbf{\Phi}_1$.
It is seen from (\ref{eq:Primal opt}) that when $\phi_{k}<1$ and $w_{k}=1$,
the $k$-th data stream will be unused. As a result, without loss of optimality,
we can set the maximum rate precoder as
\bea\label{eq:primal opt max-rate}
\Fv_{\text{max-rate}}=\Mv^{-1/2}\Vv_1(\Iv_{N_I}-\mathbf{\Phi}_1^{-1})_+^{1/2}.
\eea

\subsubsection{Quality of Service (QoS) Design}\label{sec:QoS design}
For any given weight factors, the precoder in (\ref{eq:Primal opt}) minimizes the weighted sum-MSE, i.e., $\trace(\Wv\Cv_L)$ in (P-7).
Therefore, besides the rate maximization, the solution can be exploited for handling
the error performance of each data stream so as to ensure the QoS.

Specifically, the QoS design can be achieved by setting $\Wv=\Iv_{N_I}$ and
applying the unitary discrete Fourier transform (DFT) matrix $\Dv\in\mathbb{C}^{N_I\times N_I}$
to (\ref{eq:Primal opt}) as
\bea\label{eq:primal opt QoS}
\Fv_{\text{QoS}}=\Mv^{-1/2}\Vv_1(\mathbf{\Phi}_1^{-1/2}-\mathbf{\Phi}_1^{-1})_+^{1/2}\Dv.
\eea
Here, we notify that the DFT matrix $\Dv$ enables all the MSEs, i.e., the diagonal elements of the MSE matrix $\Cv_L$ in (P-7),
have the same value without changing their sum \cite{Palomar:03} \cite{CSong:16TWC}.
Since we have $\Wv=\Iv_{N_I}$, the resulting solution minimizes the maximum MSE among data streams while maintaining the minimum sum-MSE, thereby achieving the QoS.
The QoS design is particularly useful when independent messages are spatially multiplexed across the sub-channels
and should be separately decoded.
Note that for the case of a singe S-ID user with a single antenna, i.e., $N_I=K_I=1$, the two solutions in (\ref{eq:primal opt max-rate}) and (\ref{eq:primal opt QoS}) are merged into one.
Further, if $N_P=N_E=1$, they reduce to the CR-SWIPT beamforming scheme in [15].

\subsection{Dual Variable Optimization}
The remaining problem is to determine the dual variables $\tilde{\uv}$ in (\ref{eq:Primal opt}).
Let us consider the dual problem constrained by the dual feasibility $\delta_{\min}(\Mv)>0$ (see (\ref{Apxeq:dual function joint}) in Appendix \ref{Appendix B}) as
\bea
\text{(P-8)}~~~~\sup_{\tilde{\mathbf{u}}}l(\tilde{\uv})~~s.t.~~\delta_{\min}(\Mv)>0.\nonumber
\eea
Then, following the same argument in Proposition \ref{Prop:Prop2},
we can show that the strong duality holds between (P-7) and its dual (P-8).
Therefore, the optimal dual variables in (\ref{eq:Primal opt}) can be attained by solving (P-8), which is accomplished by applying the ellipsoid method, for which
the subgradient of $l(\tilde{\uv})$ at a feasible point $\tilde{\uv}$ is computed by $[-\trace(\Fv^{\star\mathsf{H}}\Fv^\star)+P_T,\{\trace(\Fv^{\star\mathsf{H}}\Gv_{i}^\mathsf{H}\Gv_i\Fv^\star)-E_{\text{th},i}\}_{i=1}^{K_E},\{-\trace(\Fv^{\star\mathsf{H}}\Tv_j^\mathsf{H}\Tv_j\Fv^\star)+I_{\text{th},i}\}_{j=1}^{K_P}]$.
Otherwise if $\tilde{\uv}$ is infeasible, i.e., $\delta_{\min}(\Mv)\le0$,
we update $\tilde{\uv}$ utilizing the subgradient of $\delta_{\min}(\Mv)$ as
$[-1,\{\mv^{\mathsf{H}}\Gv_i^\mathsf{H}\Gv_i\mv\}_{i=1}^{K_E},\{-\mv^\mathsf{H}\Tv_i^\mathsf{H}\Tv_i\mv\}_{j=1}^{K_P}]$,
where $\mv\in\mathbb{C}^{M\times1}$ denotes the eigenvector of $\Mv$ corresponding to $\delta_{\min}(\Mv)$.
The ellipsoid updating procedure is summarized below.
The algorithm finds the global optimal solution of (P-7) attributed to the strong duality of (P-7) and (P-8) as proved in Proposition \ref{Prop:Prop2},
the primal optimal solution in Proposition \ref{Prop:Prop3}, and the convexity of (P-8) for which the ellipsoid algorithm converges to the dual optimum \cite{SBoyd:18}.

\begin{algorithm}
\caption{Joint Optimal Design for $K_I=1$.}
\begin{algorithmic}
\STATE Initialize $\tilde{\uv}\succeq\mathbf{0}$.
\REPEAT
    \STATE Compute $\Mv$ for a given $\tilde{\uv}$.
    \IF {$\delta_{\min}(\Mv)>0$}
        \STATE Set $\Wv$ and $\Dv$ as in Section \ref{sec:Max-rate Design SuMIMO} or -B2.
        \STATE Compute the primal optimal $\Fv^\star$ in (\ref{eq:Primal opt}).
        \STATE Compute the subgradient of $l(\tilde{\uv})$.
    \ELSE
        \STATE Compute the subgradient of $\delta_{\min}(\Mv)$.
    \ENDIF
    \STATE Update $\tilde{\uv}$ using the ellipsoid method subject to $\tilde{\uv}\succeq\mathbf{0}$.
\UNTIL {$\tilde{\uv}$ converges to the prescribed accuracy}.
\STATE Set $\Fv=\Fv^\star$.
\end{algorithmic}
\label{algorithm:algorithm2}
\end{algorithm}

\section{Discussion}\label{sec:Discussion}

In this section, we provide an in-depth discussion on the proposed precoder designs from a practical implementation perspective.
First, we investigate the amount of CSIs required at each secondary node and the channel estimation procedures for it.
Then, we also quantify the computational complexity of the proposed algorithms to get more insight into the design.

\subsection{Required CSIs at Each Node}\label{sec:Required CSIs at Secondary Nodes}

In order to meet all required constraints in the CR-SWIPT networks, the S-BS must control the secondary users
with global CSIs of $\{\Hv_k,\Gv_i,\Tv_j,\forall k,i,j\}$.
Therefore, it is reasonable to assume that the precoding matrix $\Fv$ is computed at the S-BS.
In contrast, the S-ID users do not need to compute $\Fv$, because only the information of effective downlink channel $\Hv_k\Fv_k$
and noise covariance $\Rv_{n,k}$ is sufficient for the $k$-th S-ID user to decode its own message as shown in (\ref{eq:rate}).
The S-EH users require neither the precoding matrix nor the CSI, since no further receive signal processing is needed for energy harvesting.
Note that there is no required CSI at the primary nodes to do with the secondary network.

\subsection{Channel Acquisition Procedure}\label{sec:Channel Acquisition Procedure}

To achieve the potential benefits of the proposed precoder designs, an accurate channel estimation at the S-BS is essential.
Thus, for channel estimation, the time division duplexing (TDD) scheme that can exploit
the channel reciprocity between a transmitter and a receiver is a better choice than the frequency division duplexing scheme.
In this subsection, we introduce TDD based channel acquisition processes to achieve the required CSIs at each node.

First, in the beginning of each channel coherence block, the secondary users transmit orthogonal training sequences to the S-BS
to allow the S-BS to estimate the CSIs of both $\mathbf{H}_{k},\forall k$ and $\mathbf{G}_{i},\forall i$.
Note that the pilot transmission of the S-EH users is also achievable by using the energy stored in their own batteries or
the energy that has been harvested in the previous transmission frame \cite{GYang:15}.
The S-BS also estimates the P-ID user channels $\Tv_j,\forall j$ by listening to the periodic uplink pilots transmitted from the P-ID users to the primary transmitter \cite{RZhang:08,RZhang:10,KyoungJae:11b}.
With the acquired CSIs, the S-BS is now able to compute the precoding matrix $\Fv$ through Algorithm 1 or 2.

Next, during the downlink training phase, the $k$-th S-ID can estimate the effective downlink channel $\Hv_k\Fv_k$
by utilizing the precoded training sequences at the S-BS. For example, the demodulation reference signaling in long-term evolution advanced (LTE-A) can be employed \cite{Dahlman:14}.
Then, the S-BS feedforwards the effective noise covariance $\Rv_{n,k}$ to each $k$-th S-ID user through the downlink control channels.\footnote{It is also possible for each S-ID user to apply the blind noise estimation scheme \cite{MFrikel:00} to estimate its own effective noise.}

\subsection{Complexity Analysis}\label{sec:Complexity Analysis}

\linespread{1.2}
\begin{table}
\centering \caption{Computational Complexity}
\begin{tabular}{|c||c|c|}
\hline
 &  Algorithm 1   & Algorithm 2\\
\hline
\hline
$\Rv_{n}$  & $\mathcal{O}\big(K_I^2N_I^3+K_I^2N_I^2M\big)$ &   - \\
\hline
\multirow{2}*{$\gamma$}  & $\mathcal{O}\big(K_IN_IM^2+K_PN_PK_IN_IM$ & \multirow{2}*{-}\\
& $+K_PN_PK_I^2N_I^2\big)$ &  \\
\hline
$\Lv$  & $\mathcal{O}\big(K_IN_I^3+K_IMN_I^2\big)$ & - \\
\hline
$\Wv$  & $\mathcal{O}\big(K_IN_I^3\big)$ & - \\
\hline\hline
\multirow{2}*{$\Kv$}  & $\mathcal{O}\big((K_EN_E+K_PN_P$ & \multirow{2}*{-} \\
& $+K_IN_I)M^2+K_I^2N_I^2M\big)$ & \\
\hline
\multirow{2}*{$\Mv$}  & \multirow{2}*{-} & $\mathcal{O}\big((K_EN_E$ \\
 & & $+K_PN_P)M^2\big)$ \\
\hline
$\Fv^\star$  & $\mathcal{O}\left(M^3+K_I^3N_I^3\right)$ & $\mathcal{O}\left(M^3+N_IM^2\right)$ \\
\hline
$I_{\text{elp}}$ \cite{SBoyd:18}  & $\mathcal{O}\left(K_E^2+K_P^2\right)$ & $\mathcal{O}\left(K_E^2+K_P^2\right)$ \\
\hline
\hline
\multirow{2}*{Total} & $N_GI_{\text{alt}}\big\{\mathcal{C}(\Rv_n,\gamma,\Lv,\Wv)$  & \multirow{2}*{$I_{\text{elp}}\mathcal{C}(\Fv^\star,\Mv)$} \\
&  $+I_{\text{elp}}\mathcal{C}(\Fv^\star,\Kv)\big\}$  &  \\
\hline
\end{tabular}
\label{table:complexity}
\end{table}
\linespread{1.0}

In what follows, we briefly examine the computational complexity of the proposed algorithms.
As it is hard to measure the exact amount of computations, we instead calculate the order of floating point operations required to find the optimal precoder $\Fv^\star$
at the S-BS.
Based on the analysis of matrix computation complexity in \cite{Golub:96}, the computational complexity of the proposed algorithms is analyzed in Table III.

The result of complexity analysis is summarized in Table \ref{table:complexity}.
Here, $\mathcal{C}(\{\mathcal{X}\})$ represents the required complexity for computing operation set $\{\mathcal{X}\}$,
and $I_{\text{elp}}$ and $I_{\text{alt}}$ denote the required number of iterations for the ellipsoid
and alternating optimization process, respectively.

First, the result confirms that the SuMIMO design in Algorithm \ref{algorithm:algorithm2} indeed obtains complexity advantage over the MuMIMO design in Algorithm \ref{algorithm:algorithm1}
for the case of $K_I=1$, because the multiple initial points and the alternating optimization process are unnecessary, not to mention the additional efforts for computing
the auxiliary filter matrices, i.e., $\mathcal{C}(\Rv_n,\gamma,\Lv,\Wv)$.
One interesting observation is that the amount of computations for Algorithms \ref{algorithm:algorithm1} and \ref{algorithm:algorithm2} increases in the orders of $N_I^3$ and $N_I$, respectively.
This means that the complexity gain of Algorithm \ref{algorithm:algorithm2} will be more pronounced as the S-ID user antenna $N_I$ grows,

However, it should be noted that as $M$ and $K_I$ increase, the MuMIMO design that can support multiple S-ID users at the same time and frequency
attains a significant throughput gain over the SuMIMO design based on the scheduling.
Therefore, Algorithm \ref{algorithm:algorithm1} is also important for achieving high data throughput in multiuser scenarios.
A careful examination on Table \ref{table:complexity} reveals that the entire complexity for Algorithm \ref{algorithm:algorithm1} is mostly influenced
by the number of antennas of the S-BS and the S-ID users, i.e., $M$ and $K_IN_I$ as in the conventional non-SWIPT or non-CR MuMIMO systems \cite{RZhang:10} \cite{CSong:16TWC} \cite{QShi:11}.
Therefore, computational complexity of the proposed designs is comparable with those in the conventional systems.

\section{Numerical Results}\label{sec:Numerical Results}
\vspace{-10pt}
\linespread{1.1}
\begin{table}[htp]
\centering \caption{Simulation Environments}
\begin{tabular}{|c||c|}
\hline
Parameter & Value \\
\hline\hline
Noise Power Spectral Density & $-100$ dBm/Hz\\
\hline
Signal Bandwidth & $10$ MHz \\
\hline
Energy Conversion Efficiency & $50\%$ ($\rho=0.5$)\\
\hline
Channel Model & Rayleigh pathloss model \\
\hline
Pathloss Exponent & $3$ \\
\hline
User Distance from S-BS & $10$ m \\
\hline
Reference Distance & $1$ m \\
\hline
Transmit Power ($P_T$) & $10\sim20$ dBm \\
\hline\hline
$(N_E,N_P,N_E,K_P)$ & $(2,1,2,2)$\\
\hline
\end{tabular}
\label{table:system_parameters}
\end{table}
\linespread{1.0}

In this section, we demonstrate the efficiency of the proposed algorithms for MuMIMO-CR-SWIPT networks through the numerical results.
For our simulations, we have chosen the system parameters as in Table \ref{table:system_parameters}.\footnote{
We have set the parameters such that the average received signal power at each secondary user appears
in a common SNR range, e.g., $10\sim20$ dB.
However, it should be noted that the proposed scheme is universally applicable to any system parameters.}
For ease of presentation, we assume that all the S-ID, S-EH, and P-ID users are located in the same distance from the S-BS.
Then, considering the Rayleigh pathloss model with the parameters in Table VI, we can construct the channel matrices
as $\Hv_k=10^{-3/2}\Hv_{k}^{(w)},\forall k$, $\Gv_i=10^{-3/2}\Gv_i^{(w)},\forall i$, and $\Tv_j=10^{-3/2}\Tv_j^{(w)},\forall j$,
where $\Hv_{k}^{(w)}$, $\Gv_i^{(w)}$, and $\Tv_j^{(w)}$ denote random matrices whose entries
are drawn from independent and identically distributed standard complex Gaussian.
We consider the same noise power at all S-ID users being equal to $\sigma_n^2=-100 \text{~dBm/Hz}\times 10\text{~MHz}=-30$ dBm
and the same interference threshold for all P-ID users, i.e., $I_{\text{th},1}=I_{\text{th},2}=I_{\text{th}}$.\footnote{
Throughout the section, we set the interference thresholds such that they appear between $0$ and a certain positive value which is smaller than
the maximum interference level that can be met by a non-CR SWIPT design considering no CR constraints.
Similarly, the energy thresholds $E_{\text{th},1}$ and $E_{\text{th},2}$ are set to be in the feasible energy region,
while being greater than the minimum energy level that is automatically achievable by a non-SWIPT CR design considering no EH constraints.}
We use an initial value $\uv=[1,0,\ldots,0]$ for (P-3). The initial values of (P-5) and (P-8) are similarly defined.
We adopt $N_G=100$ random initial points for Algorithm \ref{algorithm:algorithm1} unless specified otherwise.
For simplicity, we set $\alpha_k=1,\forall k$ for the WSR design.

\begin{figure}
\begin{center}
\includegraphics[width=3.4in,height=2.4in]{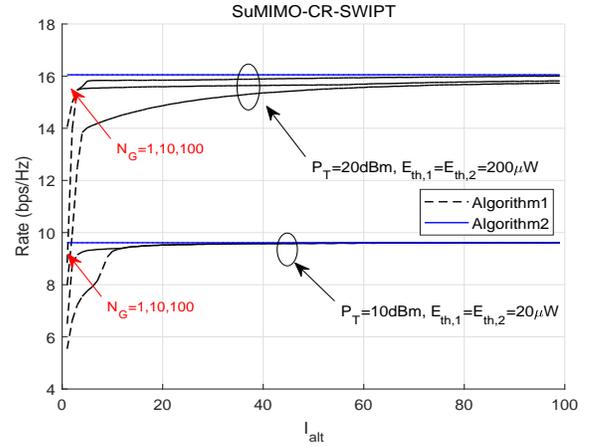}
\end{center}
\caption{Convergence trend of Algorithm 1 and 2 with $I_{\text{th}}=0.1\mu$W, $K_I=1$, and $M=N_I=4$ \label{Figure:P2P_OuterIter_Seed222}}
\end{figure}

\begin{figure}
\begin{center}
\includegraphics[width=3.4in,height=2.4in]{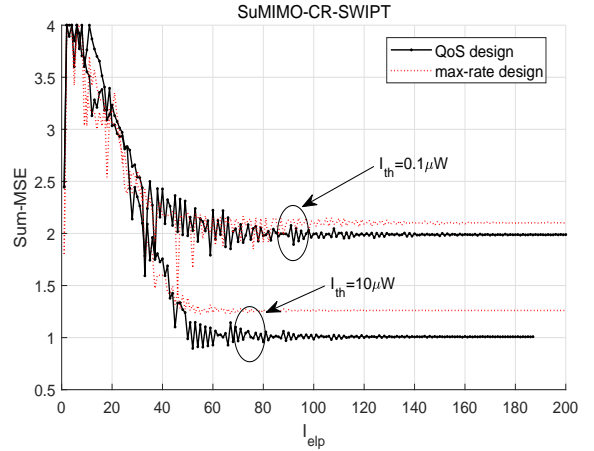}
\end{center}
\caption{Convergence trend of Algorithm 2 with $P_T=10$dBm, $E_{\text{th},1}=E_{\text{th},2}=40\mu$W, $M=4$, $K_I=1$, and $N_I=4$ \label{Figure:P2P_InnerIter_Seed222}}
\end{figure}

In Figs. \ref{Figure:P2P_OuterIter_Seed222} and \ref{Figure:P2P_InnerIter_Seed222},
we investigate the SuMIMO-CR-SWIPT networks with $K_I=1$.
Fig. \ref{Figure:P2P_OuterIter_Seed222} illustrates the convergence trend of Algorithm 1 and 2 for a system with $M=N_I=4$ and $I_{\text{th}}=0.1\mu$W.
Interestingly, we see that a few initial points may be sufficient for Algorithm 1 in $P_T=10$ dBm to achieve the maximum rate,
while a number of initial points may be needed in $P_T=20$ dBm.
The result implies that although the WMMSE cost function in (P-4) may have a convex-like form in the low SNR region,
it becomes highly non-convex as SNR goes to high.
Despite the non-convexity of (P-4), we confirm that Algorithm \ref{algorithm:algorithm2} achieves the global optimum
with a single initial point even without the alternating optimization process, and thus is efficient.

In the meantime, Fig. \ref{Figure:P2P_InnerIter_Seed222} presents the convergence trend of the ellipsoid process in Algorithm \ref{algorithm:algorithm2}
in terms of the sum-MSE performance for a system with $E_{\text{th},1}=E_{\text{th},2}=40\mu$W.
We observe that the QoS design in (\ref{eq:primal opt QoS}) achieves the minimum sum-MSE in contrast to the max-rate design.
Further, the QoS design makes all sub-channels experience the same MSE.
Therefore, the MSE gain of the QoS design leads to the bit error rate performance advantage over the max-rate design \cite{Palomar:03}.
Observe that as the interference threshold becomes tighter, the convergence speed gets slower.
The result shows that when $I_{\text{th},j}\rightarrow0^+$, the system could be inefficient,
for which the zero-interference design in Section \ref{sec:zero interference design} becomes useful.

\begin{figure}
\begin{center}
\includegraphics[width=3.4in,height=2.4in]{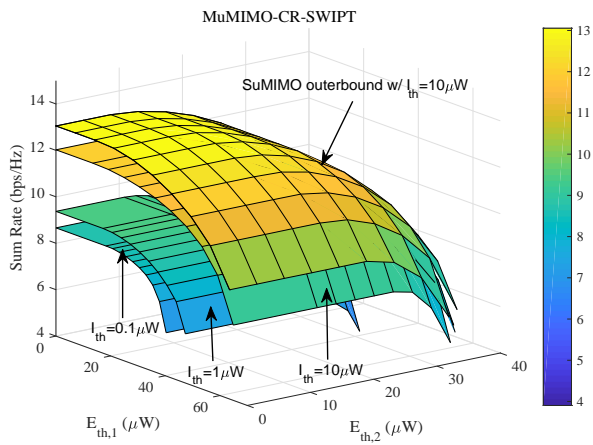}
\end{center}
\caption{Rate-energy tradeoff performance of the proposed WSR design with $P_T=10$dBm, $M=4$, and $K_I=N_I=2$ \label{Figure:Plot3D_Seed333}}
\end{figure}
\begin{figure}
\begin{center}
\includegraphics[width=3.4in,height=2.4in]{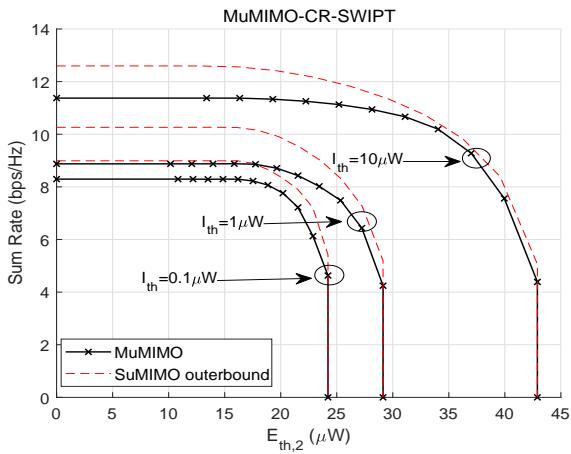}
\end{center}
\caption{Rate-energy tradeoff performance of the proposed WSR design with $P_T=10$dBm, $M=4$, $K_I=N_I=2$, and $E_{\text{th},1}=30\mu$W \label{Figure:Plot2D_Seed333}}
\end{figure}

Figs. \ref{Figure:Plot3D_Seed333} and \ref{Figure:Plot2D_Seed333} consider the MuMIMO-CR-SWIPT systems
 for a system with $K_I=N_I=2$ and $P_T=10$dBm, and show the rate-energy tradeoff performance in a 3-D plot and in its 2-D intersection at $E_{\text{th},1}=30\mu$W, respectively.
Here, we employed the WSR utility as an objective function of (P-1) to acquire the maximum WSR.
The name of {``\it SuMIMO outerbound''} amounts to the case of perfect collaboration among the multiple S-ID users, which results in a single macro S-ID user with $N_I=4$.
Unlike the case of $K_I=1$, it is generally difficult to identify whether the resulting solution is optimal or not due to the lack of knowledge on the global optimal solution.
Nevertheless, we can carefully infer that the proposed solution achieves the optimal boundary points
based on the observation that all the tradeoff regions exhibit a nice convex shape and are close to their single-user outer bounds.
At any information rate, there exists an unachievable energy region, which shows that the feasibility check in Section \ref{sec:Maximum Achievable Energy} is important before solving the problem.
Obviously, as $I_\text{th}$ decreases, the achievable tradeoff region will shrink to meet tighter CR constraints.
One interesting observation in Fig. \ref{Figure:Plot2D_Seed333} is that as the energy threshold approaches its maximum value,
the achievable rate converges to its outerbound.
This confirms our previous statement in Proposition \ref{Prop:Prop1} that
the maximum energy of an S-EH user is achievable via a single beam vector that is pointing in one direction irrespective of the S-ID user topology.

\begin{figure}
\begin{center}
\includegraphics[width=3.4in,height=2.4in]{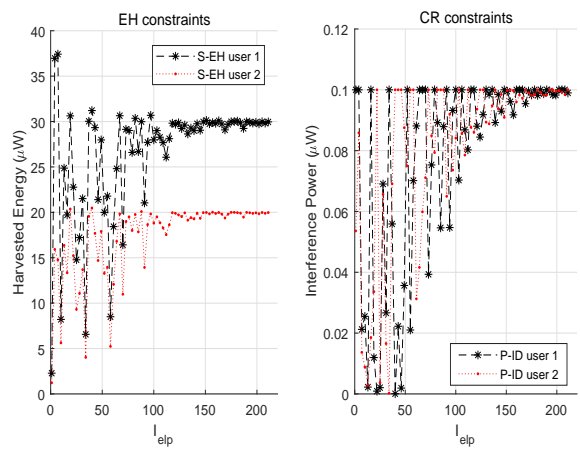}
\end{center}
\caption{Convergence trend of EH and CR constraints in MuMIMO-CR-SWIPT with $P_T=10$dBm, $M=4$, $K_I=N_I=2$, $E_{\text{th},1}=30\mu$W, $E_{\text{th},2}=20\mu$W,
and $I_{th}=0.1\mu$W \label{Figure:EH_CR_constraints_13dB_422_Seed113}}
\end{figure}

Fig. \ref{Figure:EH_CR_constraints_13dB_422_Seed113} shows a snapshot of the ellipsoid process in Algorithm 1 in terms of the harvested energy and the interference power
for a system with $M=4$ and $K_I=N_I=2$.
Here, we set the threshold values as $E_{\text{th},1}=30\mu$W, $E_{\text{th},2}=20\mu$W and $I_{\text{th},1}=I_{\text{th},2}=0.1\mu$W.
The figure confirms that the proposed algorithm achieves all the required constraints.
The interference power is kept below the threshold attributed to the power normalizing factor $\gamma$ in (\ref{eq:gamma}).

\begin{figure}
\begin{center}
\includegraphics[width=3.4in,height=2.3in]{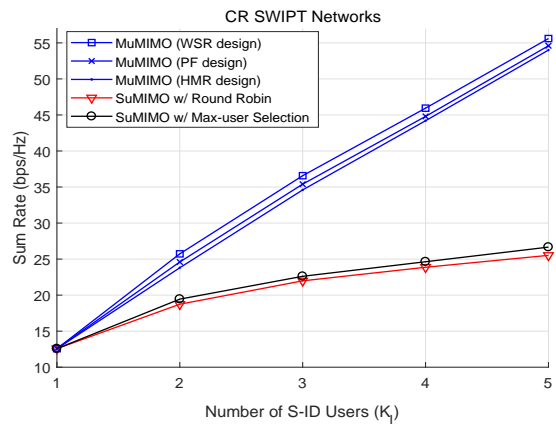}
\end{center}
\caption{WSR performance comparison with $M=K_IN_I$, $N_I=4$, $P_T=13$dBm, $E_{\text{th},1}=E_{\text{th},2}=30\mu$W,
and $I_{th}=0.1\mu$W \label{Figure:Nt_13dB_Eth3030_Ith01_Seed113}}
\end{figure}

In Fig. \ref{Figure:Nt_13dB_Eth3030_Ith01_Seed113}, we plot the WSR performance of various precoder designs in systems with
$M=N_IK_I$, $N_I=4$, and $P_T=13$ dBm.
For a fair comparison, we introduce the round robin scheduling or opportunistic max-user selection schemes \cite{AAsadi:13} to the SuMIMO design.
As expected, as $K_I$ increases, the proposed MuMIMO designs that can support multiple S-ID users simultaneously
attain significant performance advantage over the SuMIMO design based on the scheduling,
although the SuMIMO design attains complexity gain for the case of a single S-ID user.
The gain grows larger as $M$ and $K_I$ increases.

%

\begin{figure}
\begin{center}
\includegraphics[width=3.4in,height=2.4in]{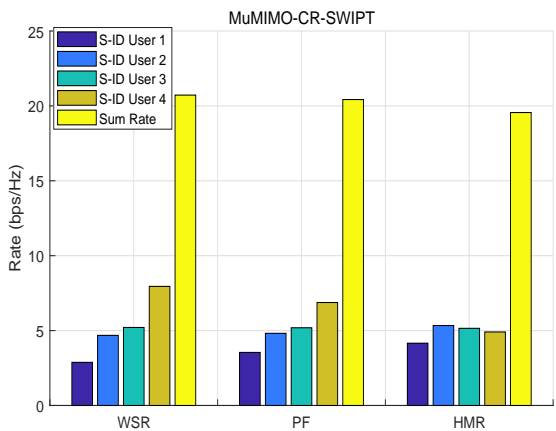}
\end{center}
\caption{Rate balancing performance of various precoder designs with $P_T=13$dBm, $M=8$, $K_I=4$, $N_I=2$, $E_{\text{th},1}=30\mu$W, $E_{\text{th},2}=20\mu$W,
and $I_{th}=0.1\mu$W \label{Figure:Bar_Fariness_Seed113}}
\end{figure}

\begin{figure}
\begin{center}
\includegraphics[width=3.4in,height=2.4in]{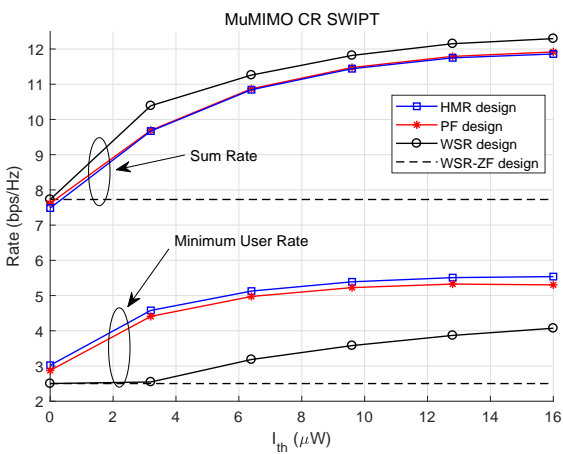}
\end{center}
\caption{WSR and minimum user rate performance of various precoder designs according to the interference thresholds $I_{th}$ with $P_T=10$dBm, $M=4$, $K_I=N_I=2$, and $E_{\text{th},1}=E_{\text{th},2}=10\mu$W \label{Figure:MinRate_Seed444}}
\end{figure}

In Fig. \ref{Figure:Bar_Fariness_Seed113}, we compare the rate balancing performance of various MuMIMO precoder designs
for a system with $M=8$, $K_I=4$, $N_I=2$, and $P_T=13$dBm.
Obviously, the WSR design achieves the best sum-rate performance.
However, the PF and HMR designs may be more attractive in terms of the worst user rate or fairness.
It is observed that the HMR design yields the best balance among the S-ID users at the cost of slight sum-rate performance loss.

Similar observation can be made in Fig. \ref{Figure:MinRate_Seed444}
which exhibits both the sum-rate and minimum user-rate performance for a system with $M=4$, $K_I=N_I=2$, and $P_T=10$dBm.
Here, ``{\it WSR-ZF design}'' denotes the WSR solution that is obtained from (P-6) with $I_{\text{th}}=0$ and $U_k(\tilde{R}_k)=\tilde{R}_k$.
We observe that although the ZF design may provide a simpler solution, it cannot achieve an additional performance gain
from both the sum-rate and minimum user rate points of view in a situation where some amount of interference is allowable.
This highlights the interference control capability of the proposed MuMIMO-CR-SWIPT designs.
As $I_{\text{th}}$ becomes smaller, the performance variation among the different utility functions diminishes.
This is because the degree of freedom of the precoder design is constrained within the null space of the P-ID user channels $\Tv$.

\section{Conclusion}\label{sec:Conclusion}

In this paper, we have investigated the optimal precoder designs for general sum-utility maximization in the MuMIMO-CR-SWIPT networks.
First, we examined the optimal energy transmission scheme to identify the feasible energy region.
Second, we proposed an efficient algorithm to find the optimal MuMIMO precoders by adopting the ellipsoid and alternating optimization process,
for which multiple initial points may be necessary to approach the global optimum.
Then, we suggested a simplified algorithm that can find a globally optimal solution without
resorting to the alternating optimization as well as the multiple initial points in a special case of a single S-ID user.
We also have offered an in-depth discussion on the proposed designs in terms of the computational complexity and the channel information requirement.
Finally, we verified the efficiency of the proposed designs via numerical simulation results.
An investigation on the tradeoff between the training phase duration and channel estimation accuracy in the imperfect CSI scenarios will be an interesting topic for future works.

\appendix

\subsection{Proof of Proposition \ref{Prop:Prop2}} \label{Appendix A}
Let $\gamma^*$, $\bar{\Fv}^*$, and $\bar{\uv}^*$ be any points that satisfy the KKT and dual feasible conditions.
Define the objective function of (P-4) as $f_0(\gamma,\bar{\Fv})=\trace(\Wv\Cv)+e(\Wv)$.
Then, by the weak duality theorem \cite{SBoyd:04}, one can show that
\bea
\label{eq:prop-1}
f_0(\gamma^*,\bar{\Fv}^*)&\geq&h(\bar{\uv}^*)\nonumber\\
&=&\inf_{\bar{\mathbf{F}}}\bar{\mathcal{L}}_{P4}(\bar{\uv}^*,\bar{\Fv})\nonumber\\
\label{eq:prop-2}
&=&\bar{\mathcal{L}}_{P4}(\bar{\uv}^*,\bar{\Fv}^*)\nonumber\\
\label{eq:prop-3}
&=&f_0(\gamma^*,\bar{\Fv}^*).\nonumber
\eea
Here, the second line follows from (\ref{eq:dual function P4}) and
the third line is due to the fact that $\bar{\mathcal{L}}_{P4}(\bar{\uv}^*,\bar{\Fv})$ is convex in
$\bar{\Fv}$ under the dual feasible condition $\Kv\succ\mathbf{0}$,
which means that the infimum of $\bar{\mathcal{L}}_{P4}(\bar{\uv}^*,\bar{\Fv})$ occurs at a point where its gradient vanishes, i.e., $\bar{\Fv}=\bar{\Fv}^*$.
In the last line, we use the complementary slackness conditions in (\ref{eq:Slack BS})-(\ref{eq:Slack CR}).
Thus, we can conclude that $f_0(\gamma^*,\bar{\Fv}^*)=h(\bar{\uv}^*)$, and the proof is completed.

\subsection{Proof of Proposition \ref{Prop:Prop3}} \label{Appendix B}

For any optimal $\Fv$ and an arbitrary unitary matrix $\Qv\in\mathbb{C}^{N_I\times N_I}$,
we can always find a modified solution $\hat{\Fv}=\Fv\Qv$ that is also optimal, since we have
\bea
&&\log\det(\hat{\Fv}^\mathsf{H}\Hv^\mathsf{H}\Rv_{n}^{-1}\Hv\hat{\Fv}+\Iv_{N_I})\nonumber\\
&=&\log\det(\Qv^\mathsf{H}\Fv^\mathsf{H}\Hv^\mathsf{H}\Rv_{n}^{-1}\Hv\Fv\Qv+\Iv_{N_I})\nonumber\\
&=&\log\det(\Fv^\mathsf{H}\Hv^\mathsf{H}\Rv_{n}^{-1}\Hv\Fv\Qv\Qv^\mathsf{H}+\Iv_{N_I})\nonumber\\
&=&\log\det(\Fv^\mathsf{H}\Hv^\mathsf{H}\Rv_{n}^{-1}\Hv\Fv+\Iv_{N_I}).\nonumber
\eea
Here, we can choose $\Qv$ such that $\hat{\Fv}^\mathsf{H}\Hv^\mathsf{H}\Rv_{n}^{-1}\Hv\hat{\Fv}$ is diagonalized without loss of optimality.
In this case, the off-diagonal elements in $\Wv$ will not affect the first term of (\ref{eq:Lagrange P4-1}).
In addition, by the Hadamard's inequality $\det(\Wv)\leq\prod_k w_{k}$,
we have $e(\Wv\big)\geq-\sum_i\log w_{k}-N_I$ with $w_{k}>0$ being the $k$-th diagoanl element of $\Wv$.
Therefore, we see that a diagonal weight matrix $\Wv$ suffices to achieve the minimum of (\ref{eq:Lagrange P4-1}).

Next, let us define a vector of the dual variables in (\ref{eq:Lagrange P4-1}) as $\tilde{\uv}\triangleq[\nu,\lambda_1,\ldots,\lambda_{K_E},\mu_1,\ldots,\mu_{K_P}]$.
Then, a dual feasible condition $\delta_{\min}(\Mv)>0$ arises for $\tilde{\uv}$, because otherwise
the corresponding dual function $l(\tilde{\uv})$ goes to $-\infty$ where
\bea\label{Apxeq:dual function joint}
l(\tilde{\uv})\triangleq\inf_{\mathbf{F}}\bar{\mathcal{L}}_{P4}.
\eea
Under the dual feasibility, $\Mv$ is invertible, which means that
one can generally express the optimal precoder as $\Fv=\Mv^{-1/2}\Vv\mathbf{\Sigma}$ for any matrix $\mathbf{\Sigma}\in\mathbb{C}^{M\times N_I}$.

Since $\Mv$ is full-rank, there are at most $N_I$ number of non-zero eigenvalues in $\mathbf{\Phi}$.
Let us further develop $\Fv$ as $\Fv=\Mv^{-1/2}\Vv_1\mathbf{\Sigma}_1+\Mv^{-1/2}\Vv_2\mathbf{\Sigma}_2$
where $\mathbf{\Sigma}_1\in\mathbb{C}^{N_I\times N_I}$ and $\mathbf{\Sigma}_2\in\mathbb{C}^{(M-N_I)\times N_I}$
are the associated sub-matrices of $\mathbf{\Sigma}=[\mathbf{\Sigma}_1^\transp~\mathbf{\Sigma}_2^\transp]^\transp$.
Then, it is true that $\Mv^{-1/2}\Vv_2\mathbf{\Sigma}_2=\mathbf{0}$ since it only increases $\bar{\mathcal{L}}_{P4}$ in (\ref{eq:Lagrange P4-1}), which leads to
\bea\label{Apxeq:primal opt 2}
\Fv=\Mv^{-1/2}\Vv_1\mathbf{\Sigma}_1.
\eea
Lastly, by substituting $\Fv$ in (\ref{eq:Lagrange P4-1}) with (\ref{Apxeq:primal opt 2}), we obtain
a modified Lagrangian as
\bea\label{Apxeq:Lagrange P4-2}
\tilde{\mathcal{L}}_{P4}=\trace\big(\Wv
(\mathbf{\Sigma}_1^\mathsf{H}\mathbf{\Phi}_1\mathbf{\Sigma}_1+\Iv_{N_I})^{-1}\big)+e(\Wv)\nonumber\\
+\trace\big(\mathbf{\Sigma}_1^\mathsf{H}\mathbf{\Sigma}_1\big)-\nu P_T+\sum_i\lambda_i E_{\text{th},i}-\sum_j\mu_j I_{\text{th},j}.
\eea

Now, we can verify from Lemma 1 and 2 below that
the Lagrangian in (\ref{Apxeq:Lagrange P4-2}) touches its minimal point when $\mathbf{\Sigma}_1$ forms a diagonal matrix because
in this case $\Wv$ in (\ref{eq:Weight matrix 2}) is given by a diagonal matrix and the first and third terms of (\ref{Apxeq:Lagrange P4-2}) can be simultaneously minimized.
Then, $\tilde{\mathcal{L}}_{P4}$ becomes convex with respect to $|\sigma_k|^2$ where $\sigma_k$ denotes the $k$-th diagonal element of $\mathbf{\Sigma}_1$.
Thus, by setting $\frac{\partial\tilde{\mathcal{L}}_{P4}}{\partial|\sigma_k|^2}=0$,
we obtain $|\sigma_k|^2=(w_k^{1/2}\phi_k^{-1/2}-\phi_k^{-1})_+$ with $\phi_k$ being the $k$-th diagonal element of $\mathbf{\Phi}_1$.
Finally, we have
\bea\label{Apxeq:Primal opt}
\Fv=\Mv^{-1/2}\Vv_1(\mathbf{W}^{1/2}\mathbf{\Phi}_1^{-1/2}-\mathbf{\Phi}_1^{-1})_+^{1/2}.\nonumber
\eea
\begin{Lemma}[\!\cite{Golub:96}]
For any square matrix $\Av$,
it is true that $\text{Tr}(\Av\Av^\mathsf{H})\geq\sum_{i}|a_i|^2$ where $a_{i}$ stands for the $i$-th diagonal element of $\Av$.
\end{Lemma}
\begin{Lemma}[\!\cite{Komaroff:90}]
For any positive definite matrix $\Bv$, we have $\text{Tr}(\Bv^{-1})\geq \sum_{i=1}^{M}b_{i}^{-1}$
where $b_{i}$ stands for the $i$-th diagonal element of $\Bv$.
\end{Lemma}
\bibliographystyle{IEEEtr}
\input{bibliography.filelist}

\end{document}